\documentclass{aa}
\usepackage{graphicx}
\usepackage{txfonts}
\usepackage{natbib}
\usepackage{amssymb,amsmath}
\bibpunct{(}{)}{;}{a}{}{,}

\usepackage[a4paper=true,pagebackref=true]{hyperref}
\hypersetup{pdftitle = The title of my PDF, pdfauthor = My name, pdfsubject= The subject, pdfkeywords = keyword1 keyword2 keyword3}
\hypersetup{colorlinks = true, linkcolor = green, anchorcolor = red, citecolor = blue, filecolor = red, pagecolor = red, urlcolor = red}

\begin{document}

\title{The Southern Twenty-centimetre All-sky Polarization Survey (STAPS): survey description and maps}

\author{
Xiaohui Sun\inst{\ref{ynu}} \and 
Marijke Haverkorn\inst{\ref{imapp}} \and 
Ettore Carretti\inst{\ref{inaf}} \and 
Tom Landecker\inst{\ref{drao}} \and 
B. M. Gaensler\inst{\ref{uc}}\and
Sergio Poppi\inst{\ref{cagliari}}\and
Lister Staveley-Smith\inst{\ref{icrar}}\and
Xuyang Gao\inst{\ref{naoc}}\and
Jinlin Han\inst{\ref{naoc}}
}

\institute{School of Physics and Astronomy, Yunnan University, Kunming 650500, China\\\email{xhsun@ynu.edu.cn}\label{ynu}
\and
Department of Astrophysics/IMAPP, Radboud University, Nijmegen, PO Box 9010, 6500 GL Nijmegen, the Netherlands\label{imapp}
\and
INAF Instituto di Radioastronomia, Via Gobetti 101, 40129 Bologna, Italy\label{inaf}
\and
National Research Council Canada, Herzberg Research Centre for Astronomy and Astrophysics, Dominion Radio Astrophysical Observatory, P.O. Box 248, Penticton, BC V2A 6J9, Canada\label{drao}
\and
Department of Astronomy and Astrophysics, University of California Santa Cruz, Santa Cruz, CA 95064, USA\label{uc}
\and
 INAF -- Osservatorio Astronomico di Cagliari, Via della Scienza 5, I-09047 Selargius, Italy\label{cagliari} 
\and
International Centre for Radio Astronomy Research (ICRAR), University of Western Australia, 35 Stirling Hwy, Crawley, WA 6009, Australia\label{icrar}
\and
National Astronomical Observatories, Chinese Academy
  of Sciences, Beijing 100101, PR China\label{naoc}
}
\date{Received xxx; accepted xxx}

\abstract{We present data processing and verification of the Southern Twenty-centimetre All-sky Polarization Survey (STAPS) conducted with Murriyang, the Parkes 64-m telescope. The survey covers the sky area of $-89\degr<{\rm Dec}<0\degr$ and the frequency range of 1.3-1.8~GHz split into 1-MHz channels. STAPS was observed commensally with the S-band Polarization All-Sky Survey (S-PASS). The survey is composed of long azimuth scans, which allows us to absolutely calibrate Stokes $Q$ and $U$ with the data processing procedure developed for S-PASS. We obtain $I$, $Q$, and $U$ maps in both flux density scale (Jy beam$^{-1}$) and main beam brightness temperature scale (K), for the 301 frequency channels with sufficiently good data. The temperature scale is tied to the Global Magneto-ionic Medium Survey (GMIMS) high-band north sky survey conducted with the Dominion Radio Astrophysical Observatory 26-m telescope. All the STAPS maps are smoothed to a common resolution of $20\arcmin$. The rms noise per channel ranges from about 16~mK to 8~mK for $I$, and from about 8~mK to 5~mK for $Q$ and $U$ at frequencies from 1.3 to 1.8~GHz. The rms noise in $Q$ and $U$ varies with declination and reaches minimum at declination of $-89\degr$. We also run rotation measure (RM) synthesis and RM clean to obtain peak polarized intensity and Faraday depth maps. The whole STAPS data processing is validated by comparing flux densities of compact sources, pixel flux density versus pixel flux density for Cen A, pixel temperature versus pixel temperature for the entire survey area, and RMs of extragalactic sources between STAPS and other measurements. The uncertainty of the flux density scale is less than 10\%. STAPS delivers an L-band ($\lambda$20~cm) multi-frequency polarization view of the Galaxy, and will help advance our understanding of the Galactic magnetic field and magnetized interstellar medium.}


\maketitle
\titlerunning{STAPS: survey description and maps}
\authorrunning{Sun et al.}

\section{Introduction}

The Galactic interstellar medium (ISM) is magnetized and contains multiple phases on various physical scales. Both magnetic fields and turbulence play crucial roles in star formation~\citep{federrath+12}, and cosmic ray acceleration and propagation~\citep{lazarian+yan+14}. The magnetic field and turbulence are coupled to each other, and a full description requires magnetohydrodynamics, which is very challenging from both observational and theoretical perspectives~\citep{cho+03, burkhart+21}. Radio polarization observations provide measurements of the magnetic field with both ordered and random components and will therefore help to understand the turbulent magnetized ISM in the Galaxy. 

The linearly polarized emission originates from synchrotron radiation of relativistic electrons spiraling around the magnetic field. When propagating through the magneto-ionic medium, the polarized wave experiences Faraday rotation, which changes polarization angle ($\chi$) as $\chi=\chi_0 + {\rm RM}\,\lambda^2$. Here $\chi_0$ is the intrinsic polarization angle, $\lambda$ is wavelength, and RM is the rotation measure defined as ${\rm RM} = 0.81 \int_{\rm source}^{\rm observer} n_e B_{\parallel}{\rm d}l$, where $n_e$ is the thermal electron density in cm$^{-3}$, $B_\parallel$ is the magnetic field along the line of sight in $\mu$G, $l$ is the path length in pc, and RM is in rad~m$^{-2}$. 

The amount of Faraday rotation differs along the line of sight and across the observation beam. This results in a complicated polarization morphology that was first observed several decades ago~\citep[e.g.][]{wieringa+93}. Images of polarized emission are thus imprinted with properties of magnetic field, high-energy electrons, and thermal electrons. Understanding these ISM ingredients requires polarization observations, which motivates polarization surveys of a large sky area.   

\begin{figure*}
    \centering
    \includegraphics[width=0.47\textwidth]{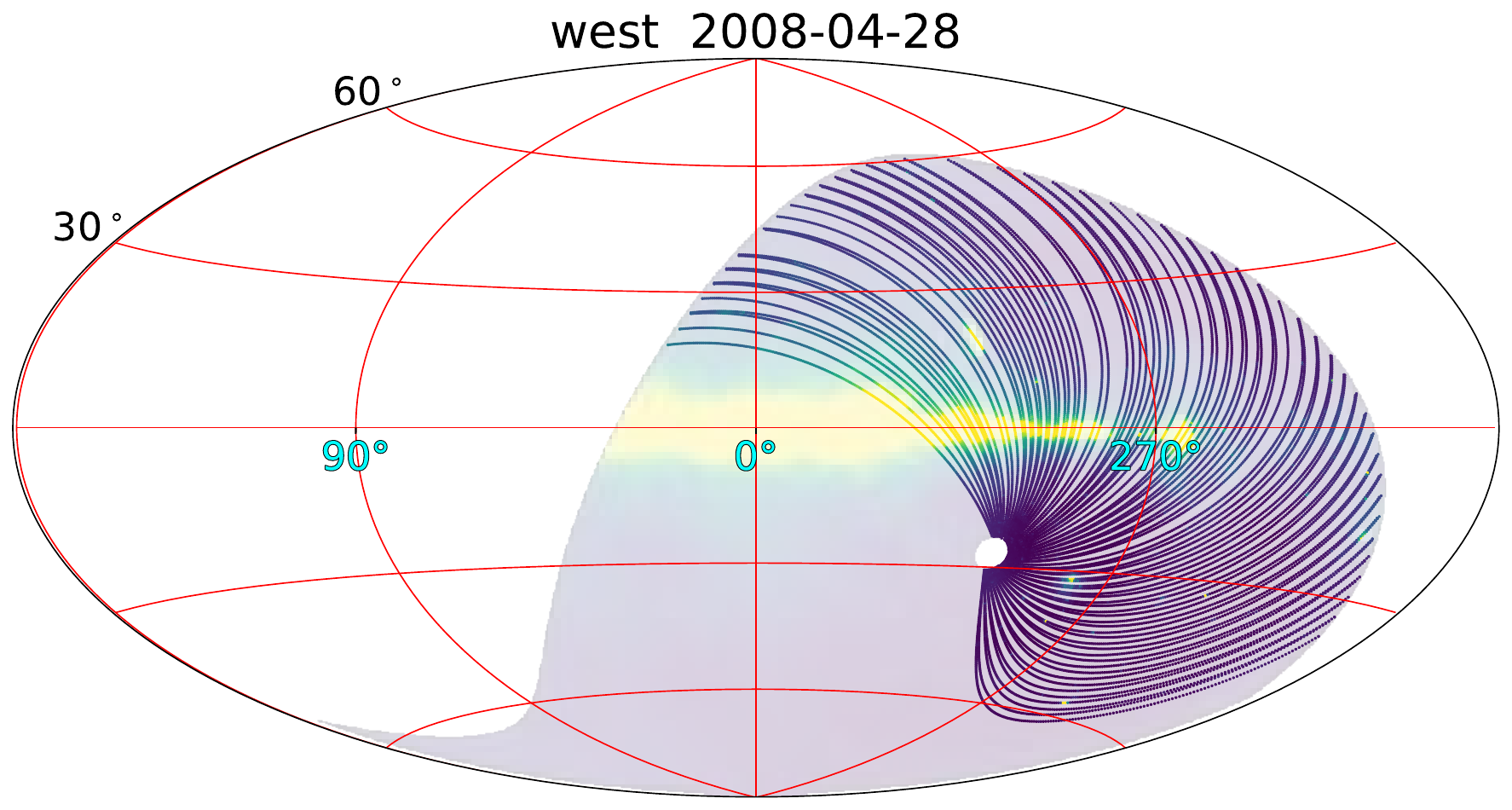}
    \includegraphics[width=0.47\textwidth]{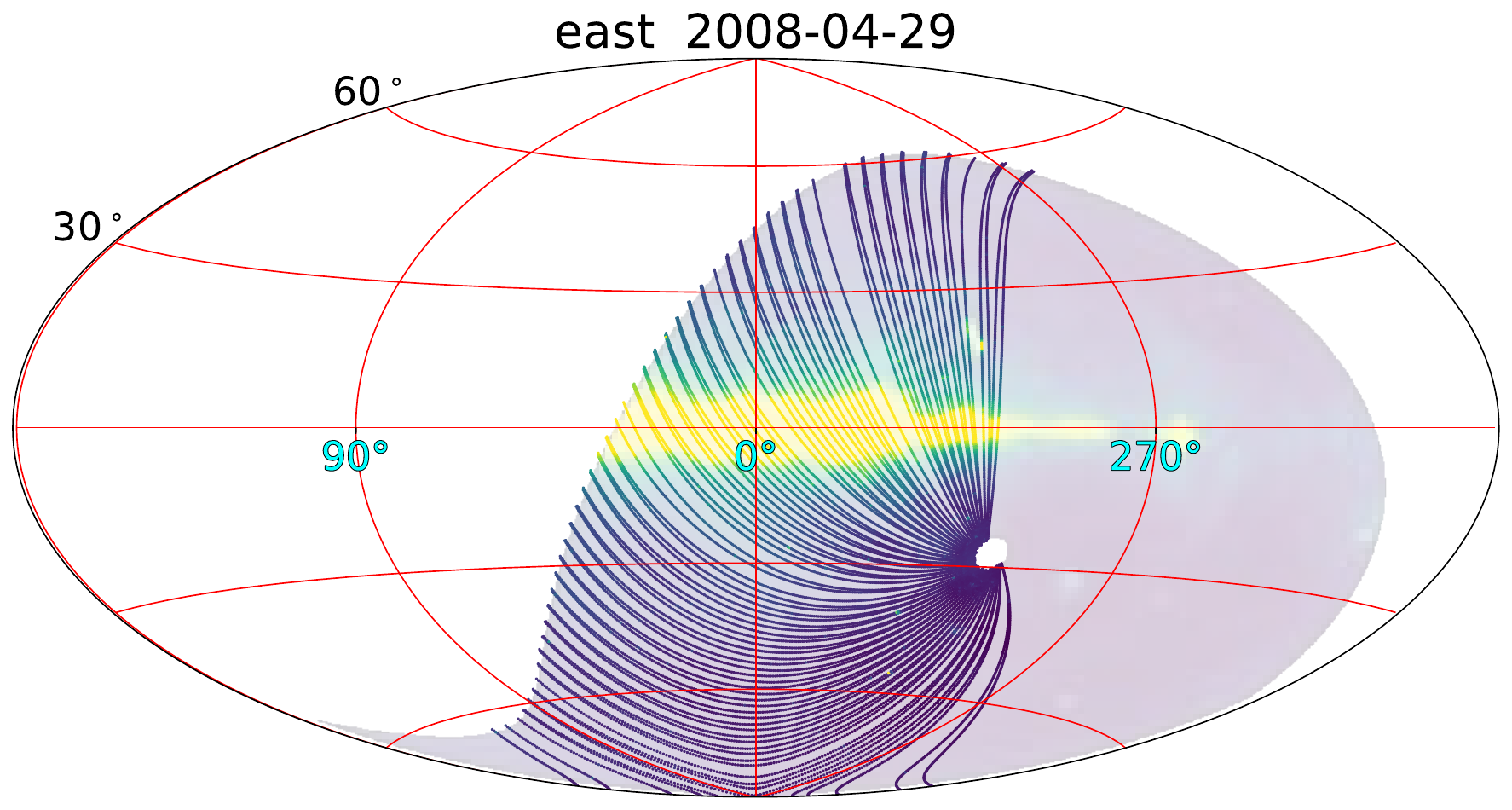}
    \includegraphics[width=0.47\textwidth]{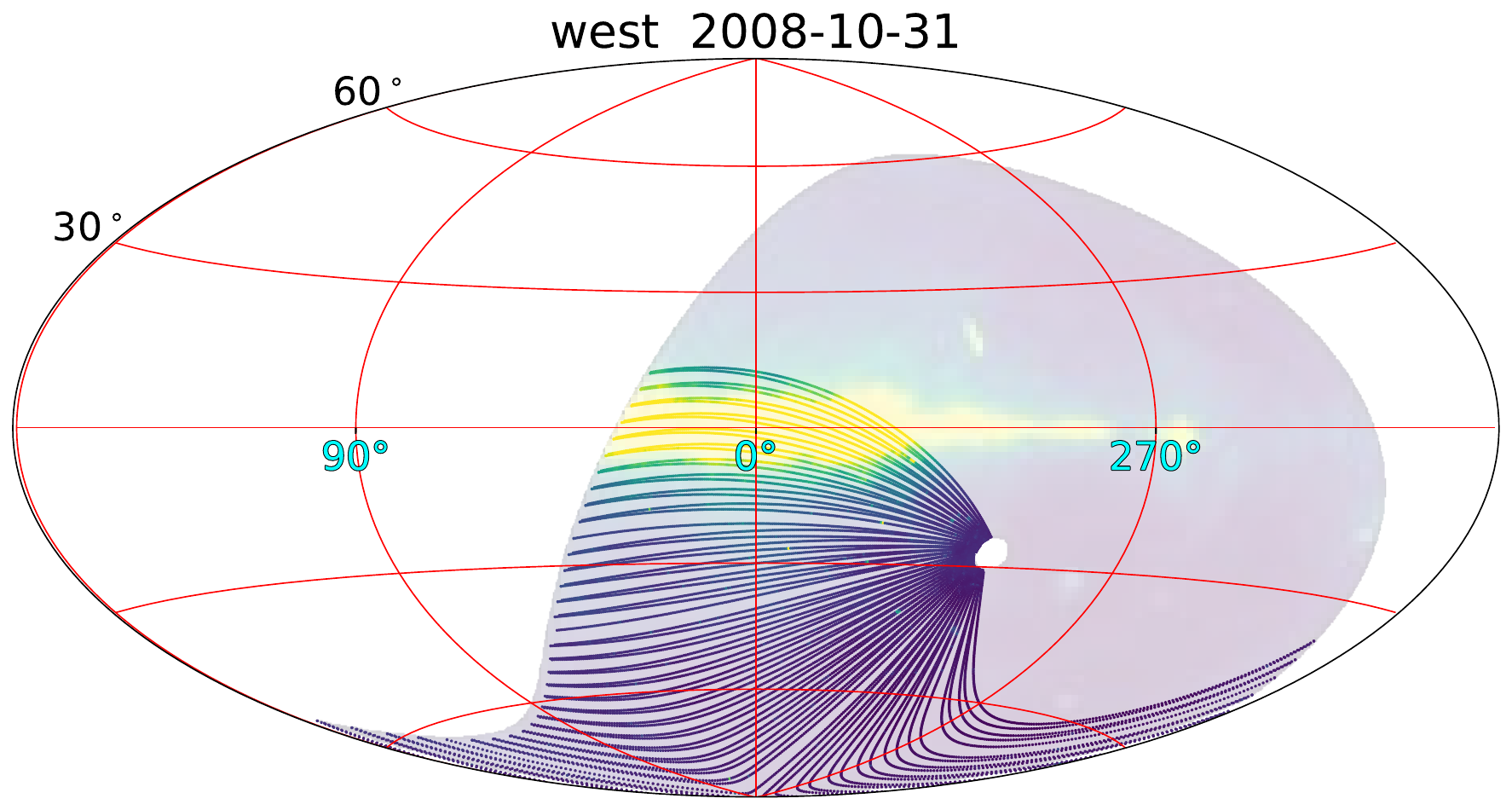}
    \includegraphics[width=0.47\textwidth]{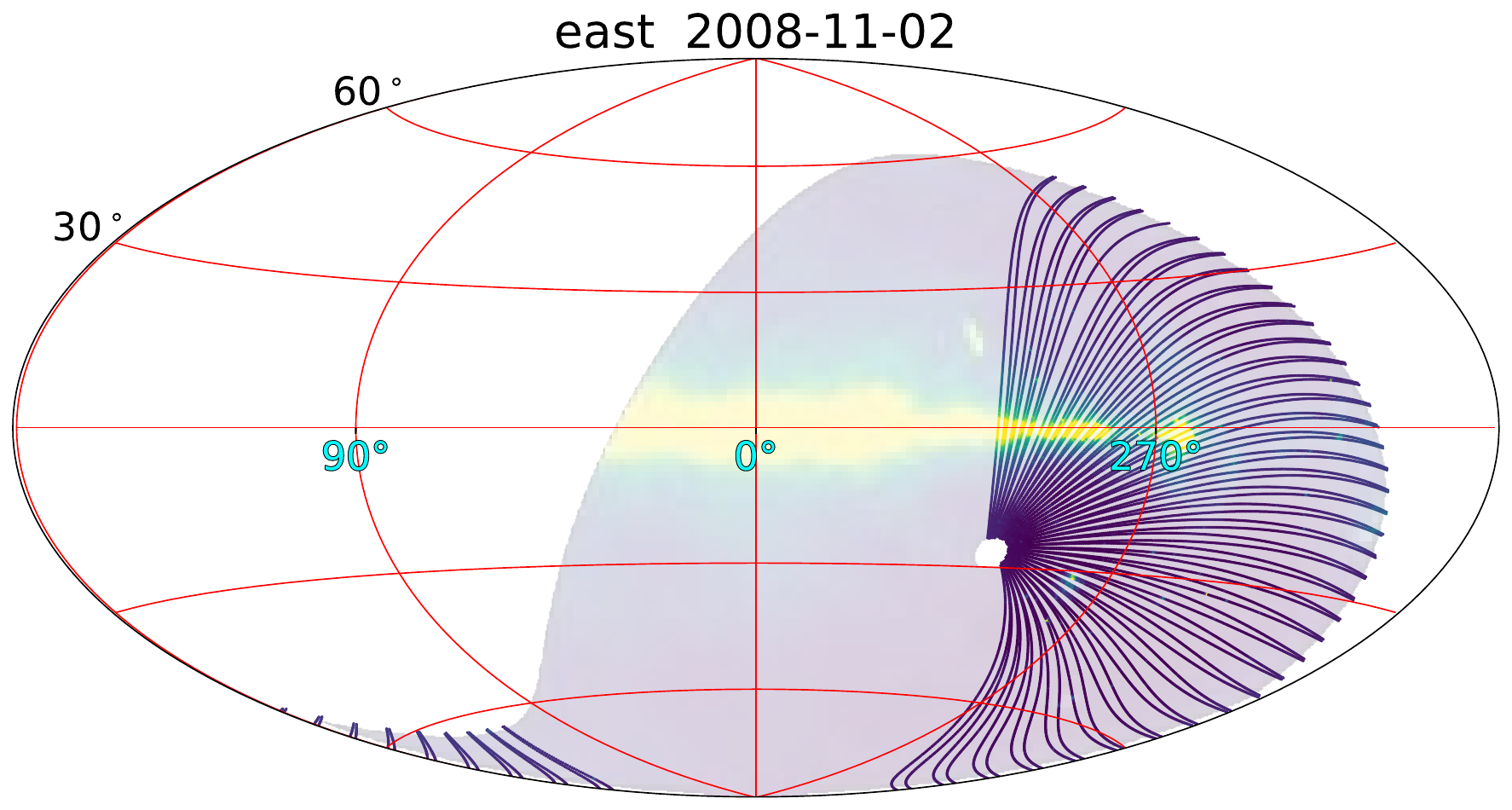}
    \caption{West and east scans conducted in April and October runs in 2008. The background image displays the total intensity at 1.4~GHz from STAPS. The images are in Galactic coordinates with Aitoff projection.}
    \label{fig:scans}
\end{figure*}

Previous low-frequency polarization surveys include the northern sky survey with the Dwingeloo 25-m telescope at frequencies 408, 465, 610, 820, and 1411~MHz~\citep{brouw+76,spoelstra84}, the northern sky survey with the Dominion Radio Astrophysical Observatory (DRAO) 26-m telescope at 1.4~GHz~\citep{wolleben+06}, and the southern sky survey with the Villa Elisa 30-m telescope at 1.4~GHz~\citep{testori+08}. The latter two have been combined to form an all-sky polarization image of the Galaxy at 1.4~GHz~\citep{reich06}. All of these surveys used narrow band receivers with a single frequency channel. 

Polarization surveys have revived over the last decade because of the development of broadband multi-channel polarization backends. In addition to the great improvement of sensitivity, the RM synthesis method~\citep{brentjens+05} can be readily employed to reconstruct the polarized intensity as a function of the Faraday depth, called the Faraday depth spectrum~\citep{sun+15b} or the Faraday dispersion function. Here, the Faraday depth is defined as
\begin{equation}
\varphi(\vec{r}) = 0.81 \int_r^{\rm observer} n_e B_{\parallel}{\rm d}l,
\end{equation}
where $\vec{r}$ is the position vector in the source, and the other quantities are the same as in the definition of RM. The Faraday depth spectrum provides information on the dominant Faraday rotation components along the line of sight and the complexity introduced by mixed emission and rotation in the magnetized ISM.

The Global Magneto-Ionic Medium Survey~\citep[GMIMS,][]{wolleben+09} is the first all-sky polarization observation campaign aiming to produce polarization images at multi-frequency channels using telescopes in both hemispheres. It consists of north and south surveys covering the approximate frequency range 300--1800 MHz. Two of the surveys have been completed with data released: the high-band north survey with the DRAO 26-m telescope over the frequency range 1280--1750~MHz split into 236.8-kHz channels~\citep{wolleben+21}, and the low-band south survey with Murriyang, the Parkes 64-m telescope over the frequency range 300--480~MHz split into 0.5-MHz channels~\citep{wolleben+19}. Based on these GMIMS surveys, individual Galactic features have been studied, such as the North Polar Spur~\citep{sun+nps+15}, the Fan region~\citep{hill+17}, the ISM towards the HII region Sh 2-27~\citep{thomson+19} and Loop II~\citep{Thomson+21}. The Galactic magnetic field has also been studied using moment maps calculated from the Faraday depth cubes~\citep{dickey+22, raycheva+24}. 

At higher frequency, the S-band Polarization All-Sky Survey~\citep[S-PASS,][]{carretti+19} observed the southern sky with Murriyang over the frequency range 2.2--2.4~GHz with channel width of 0.5~MHz. Commensal with S-PASS, the Southern Twenty-centimeter All-sky Polarization Survey~\citep[STAPS,][]{haverkorn15} covering the frequency range 1.3--1.8~GHz with channel width of 1~MHz has also been conducted as part of GMIMS, being its high-band south component survey.

This paper presents data processing and images of the total intensity and linear polarization from STAPS. The layout of the paper is as follows: the observations and data processing are briefly described in Sect.~\ref{sec:obs}; the processing is verified in Sect.~\ref{sec:verification}; the Stokes parameters $I$, $Q$, $U$ and the RM maps are presented in Sect.~\ref{sec:maps}; a summary of the results is presented in Sect.~\ref{sec:con}.

\begin{figure*}
    \centering
    \includegraphics[width=0.32\textwidth]{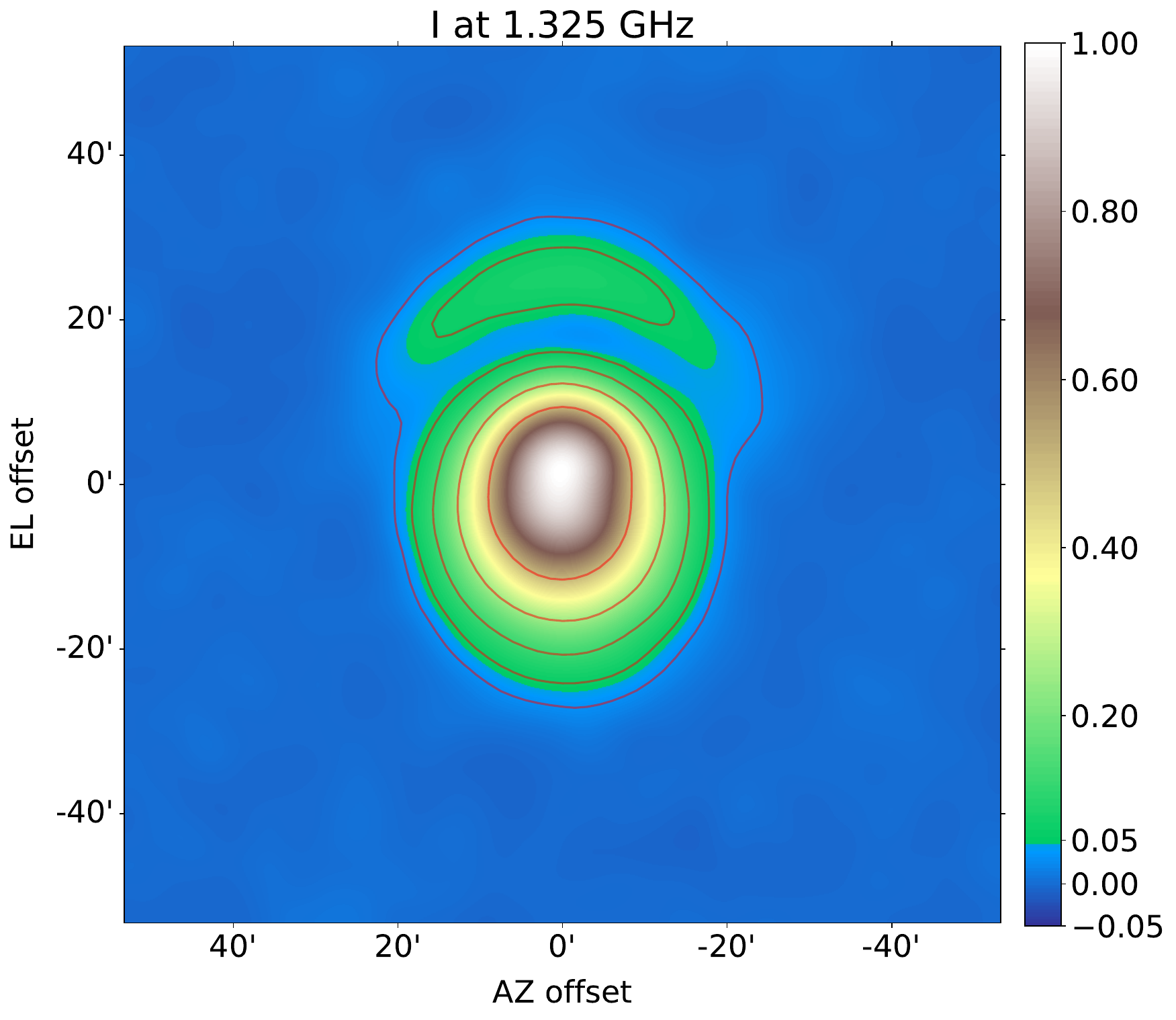}
    \includegraphics[width=0.32\textwidth]{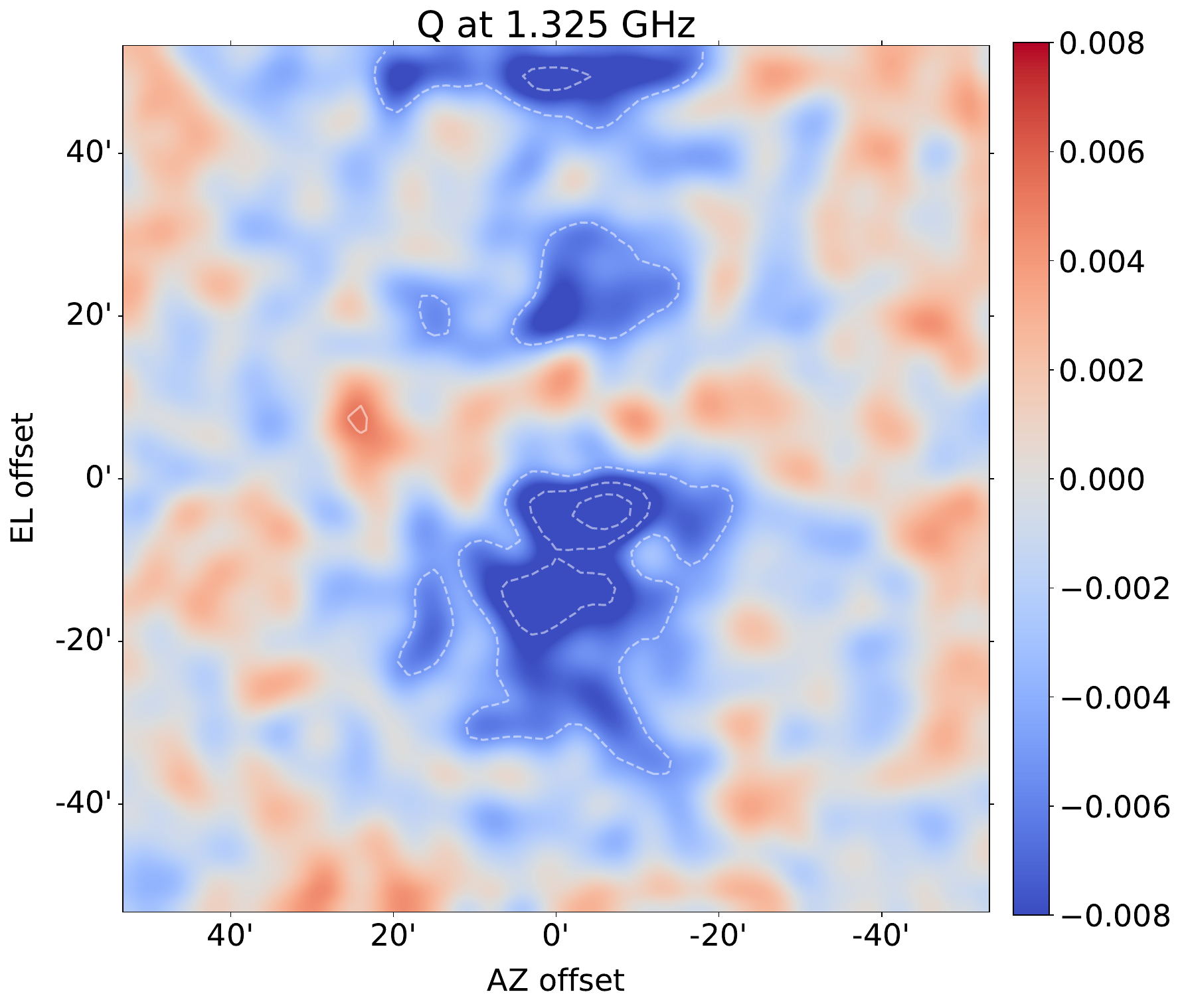}
    \includegraphics[width=0.32\textwidth]{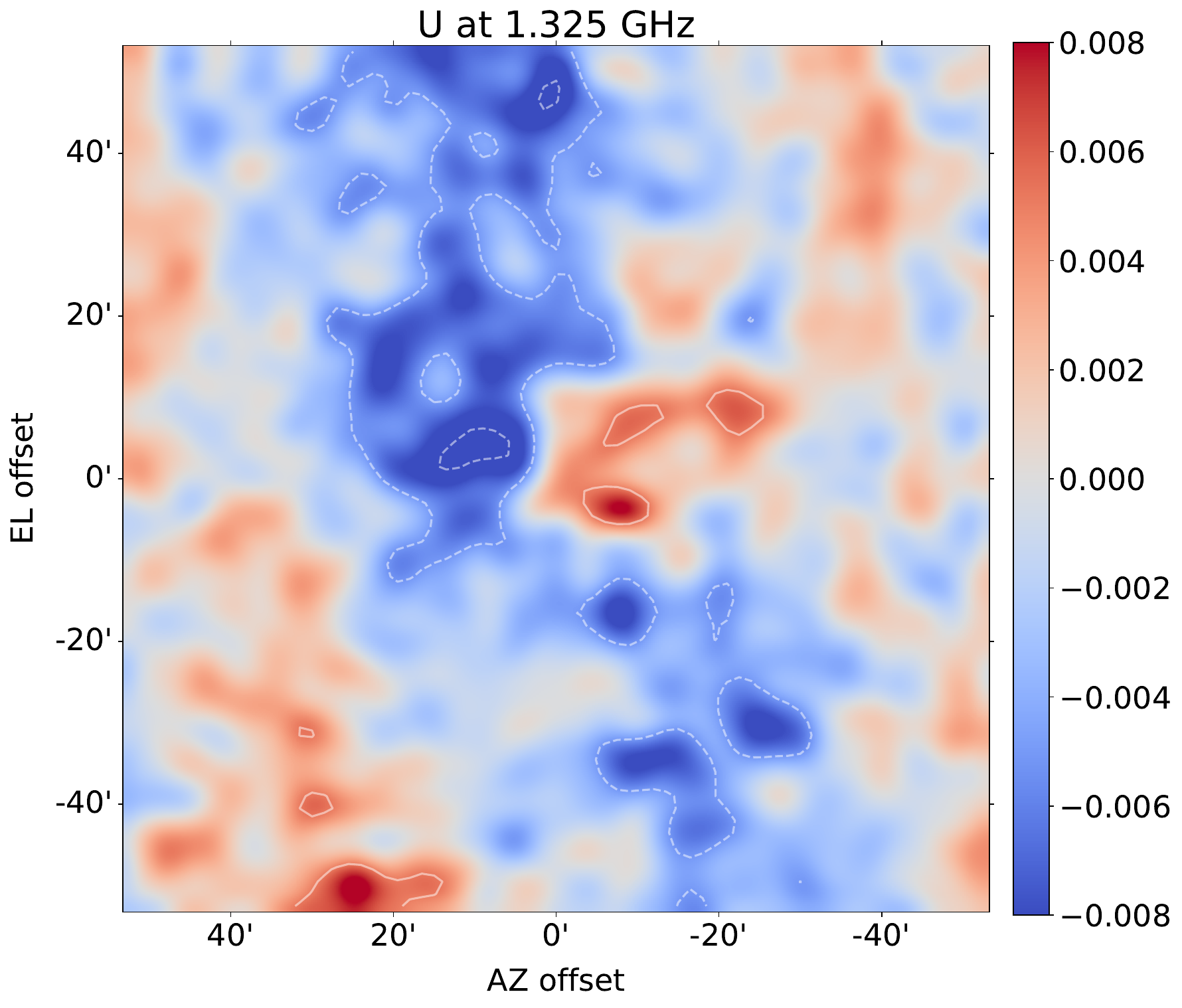}
    \includegraphics[width=0.32\textwidth]{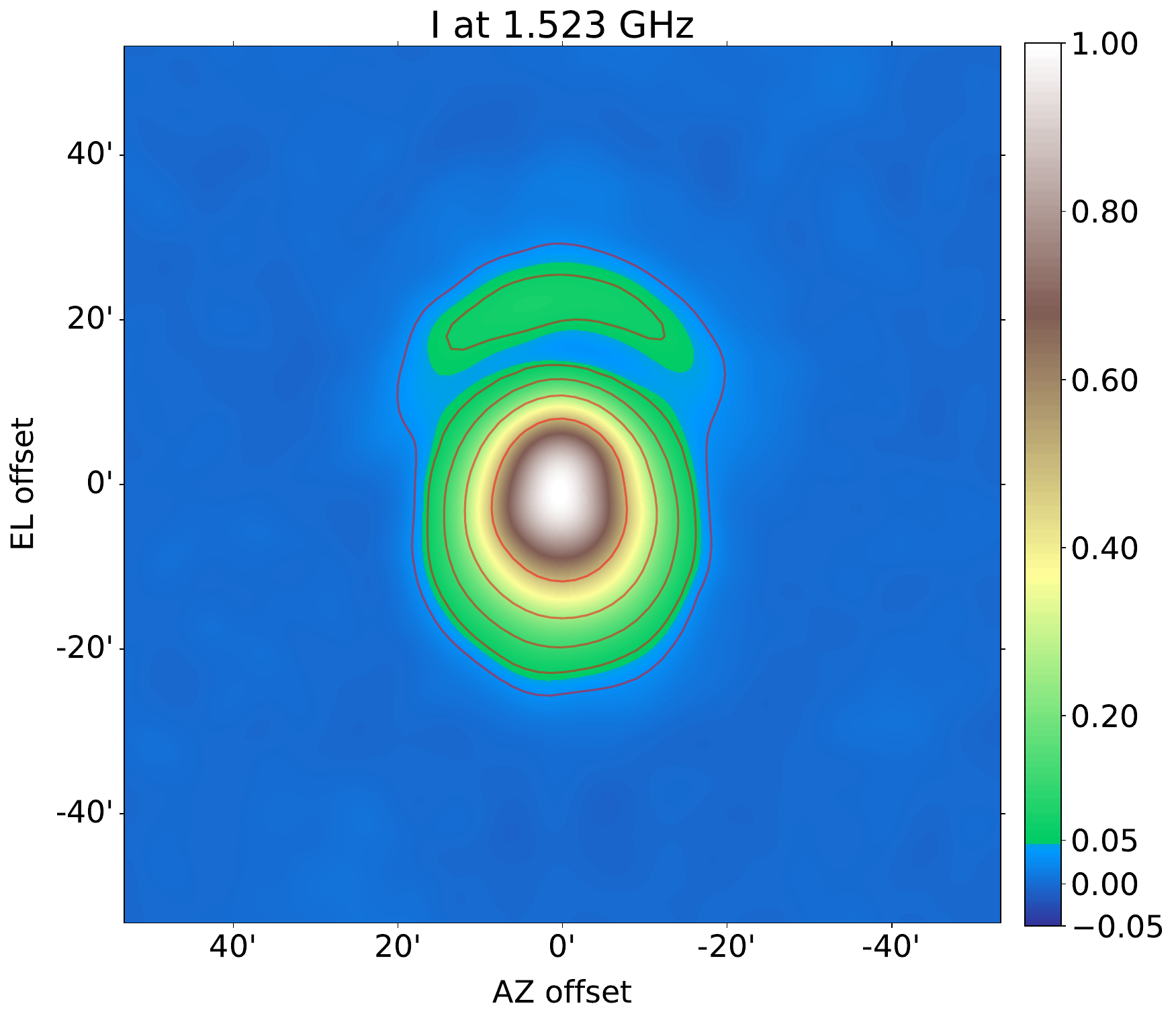}
    \includegraphics[width=0.32\textwidth]{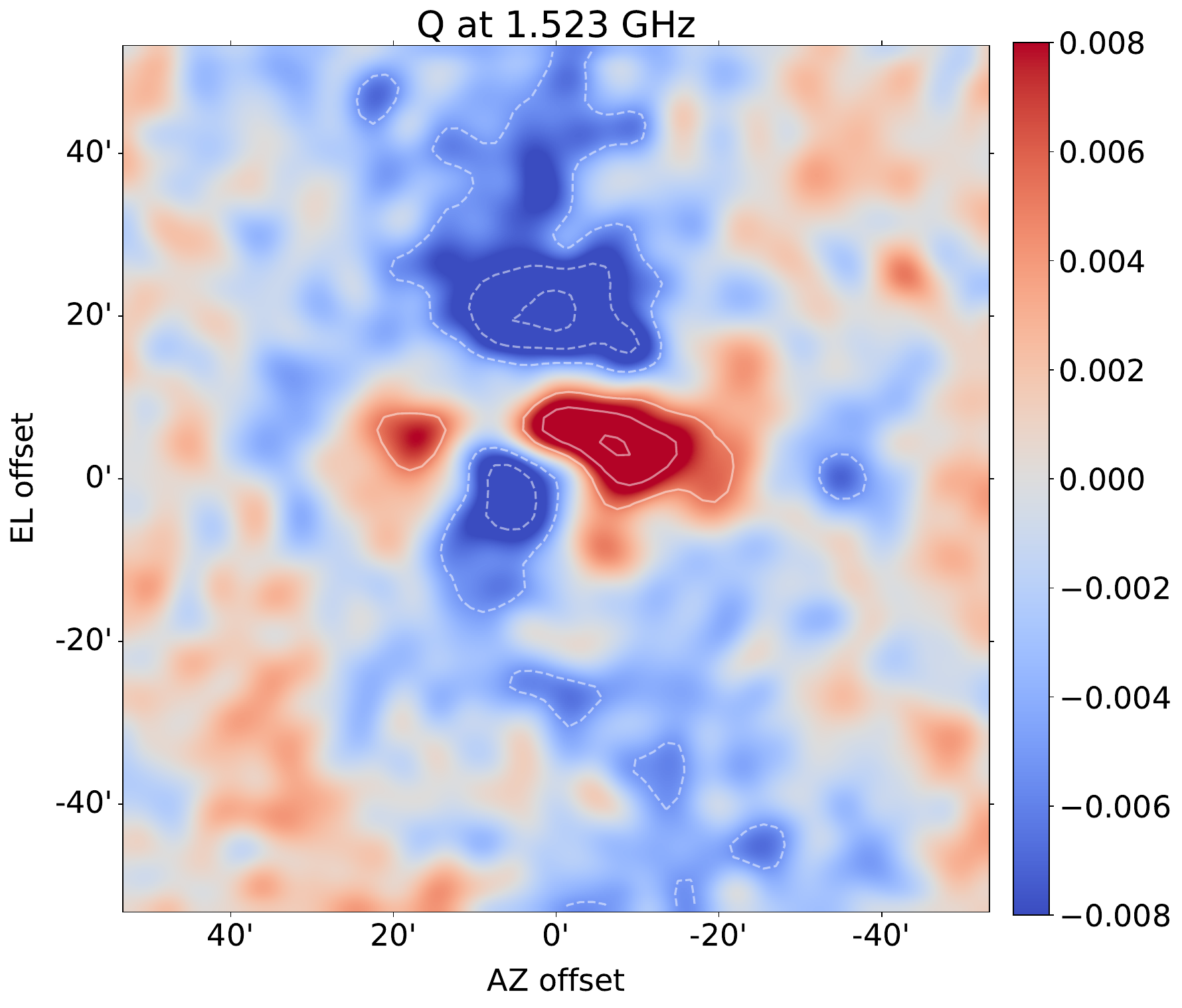}
    \includegraphics[width=0.32\textwidth]{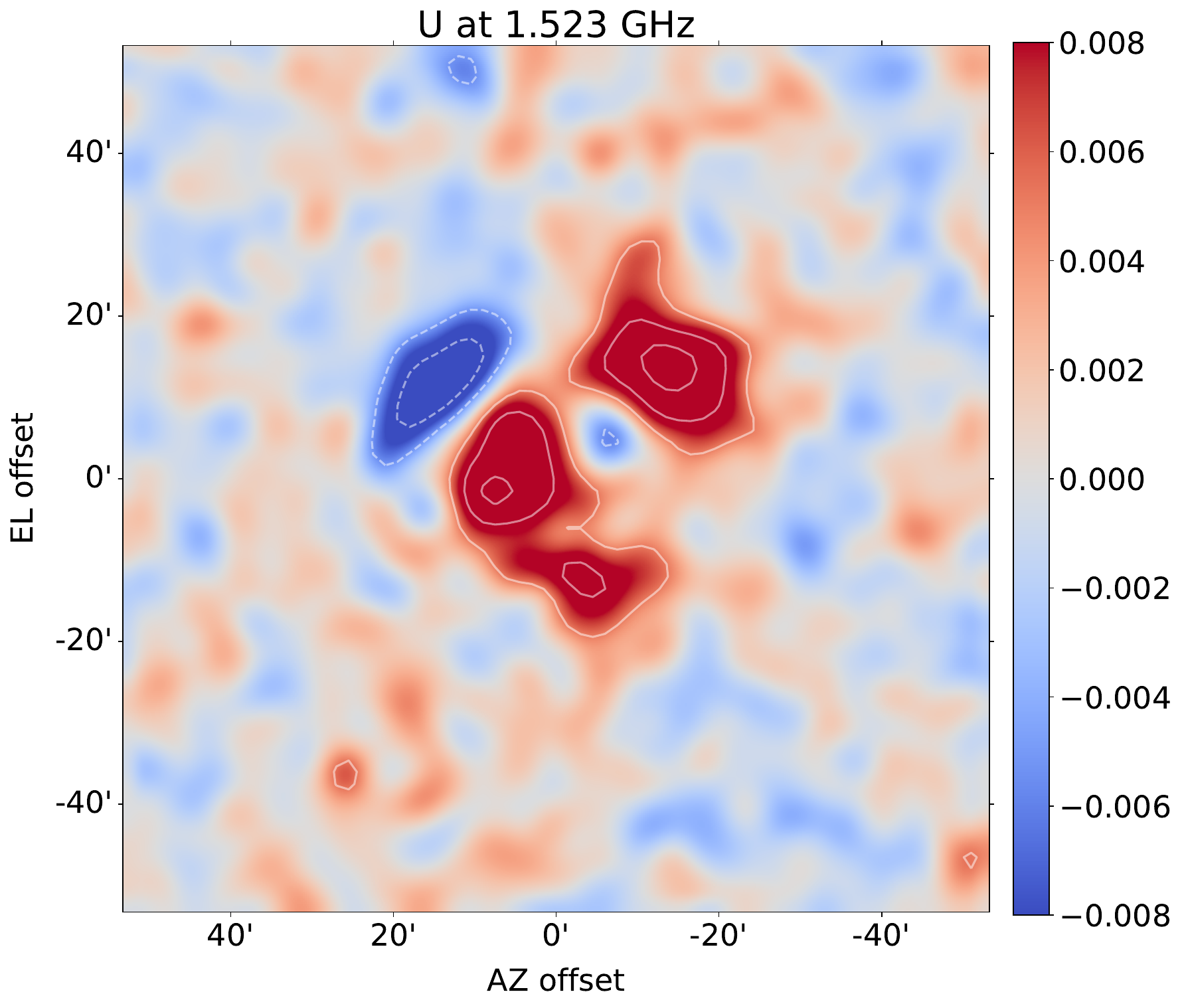}
    \includegraphics[width=0.32\textwidth]{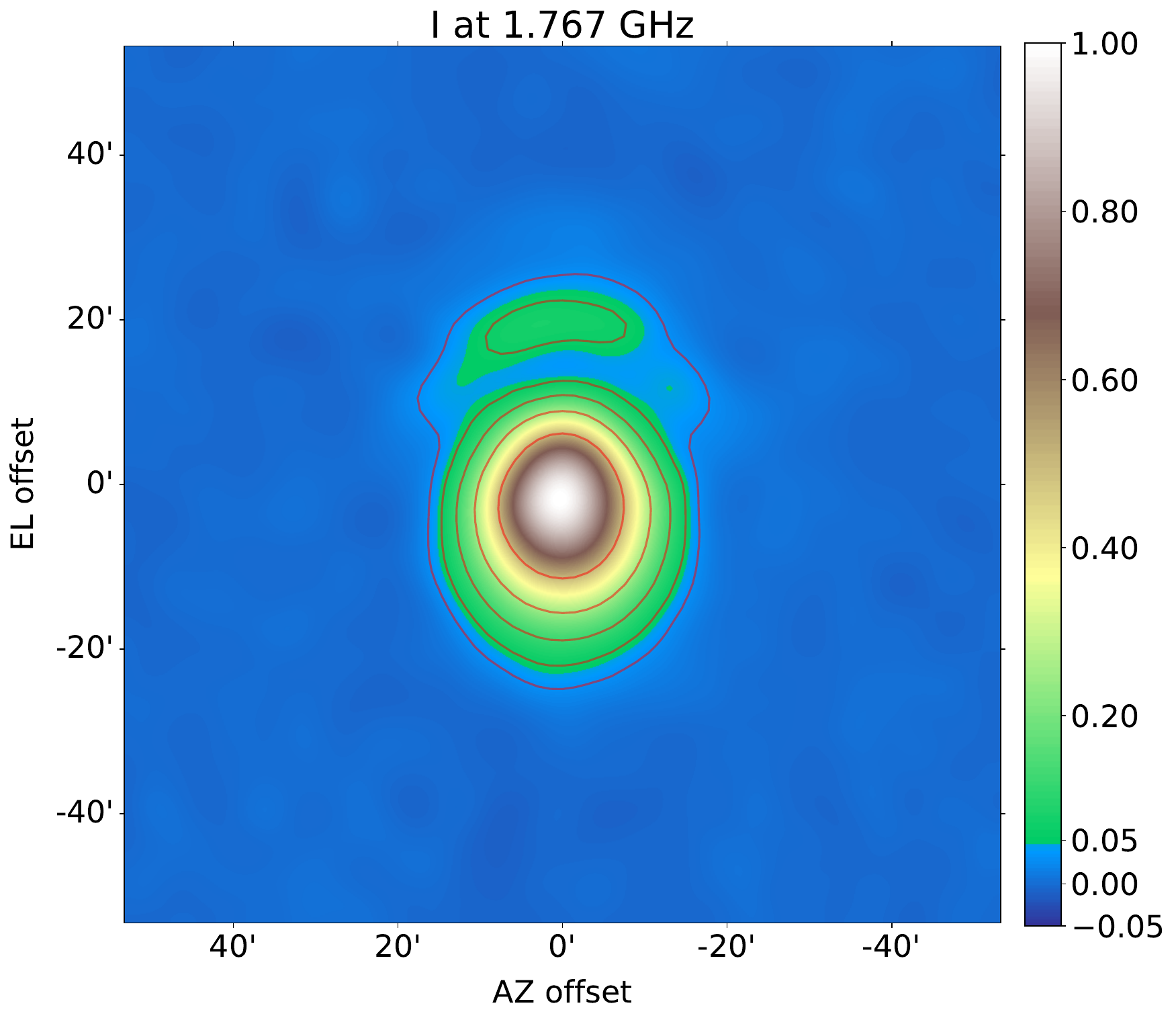}
    \includegraphics[width=0.32\textwidth]{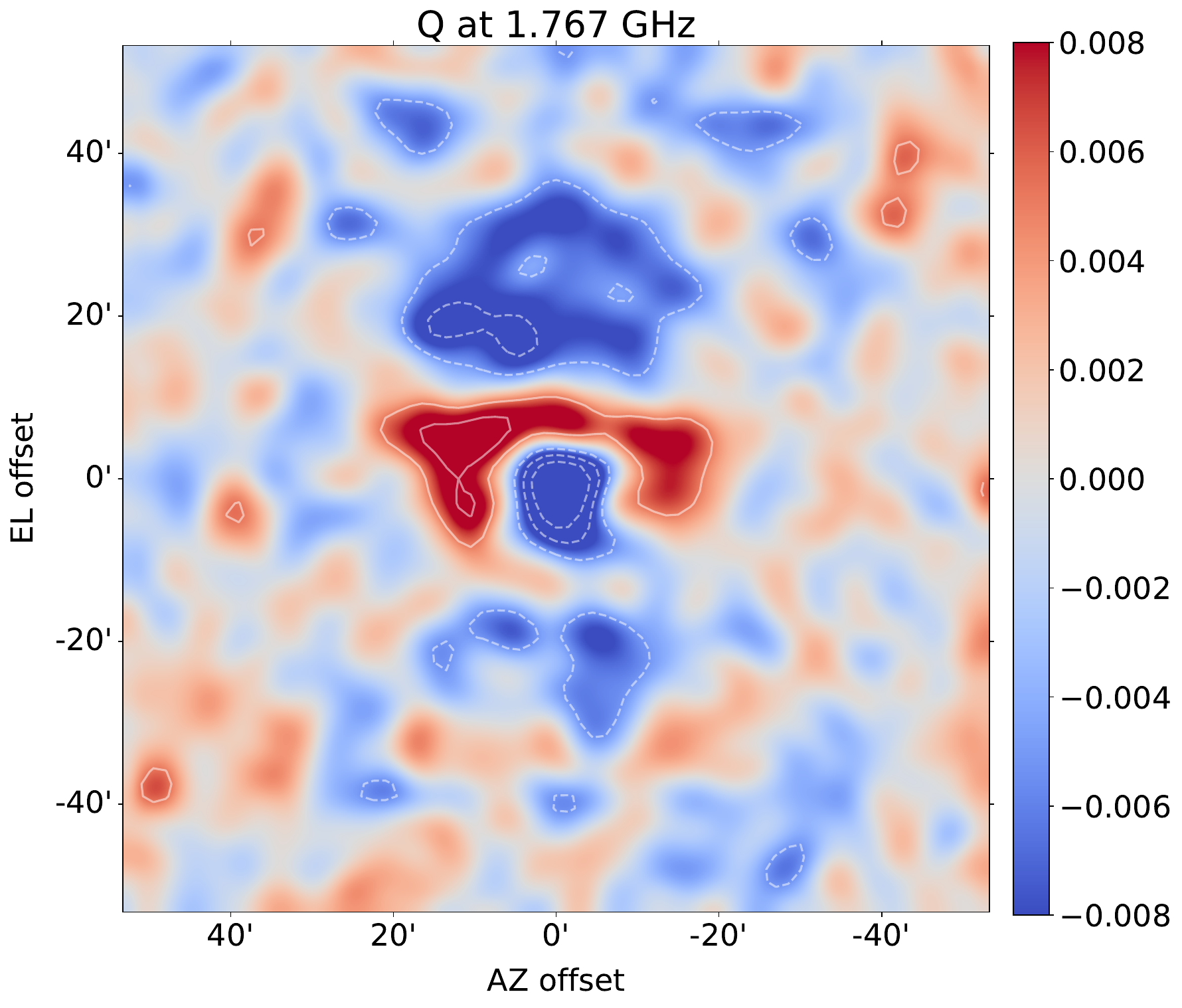}
    \includegraphics[width=0.32\textwidth]{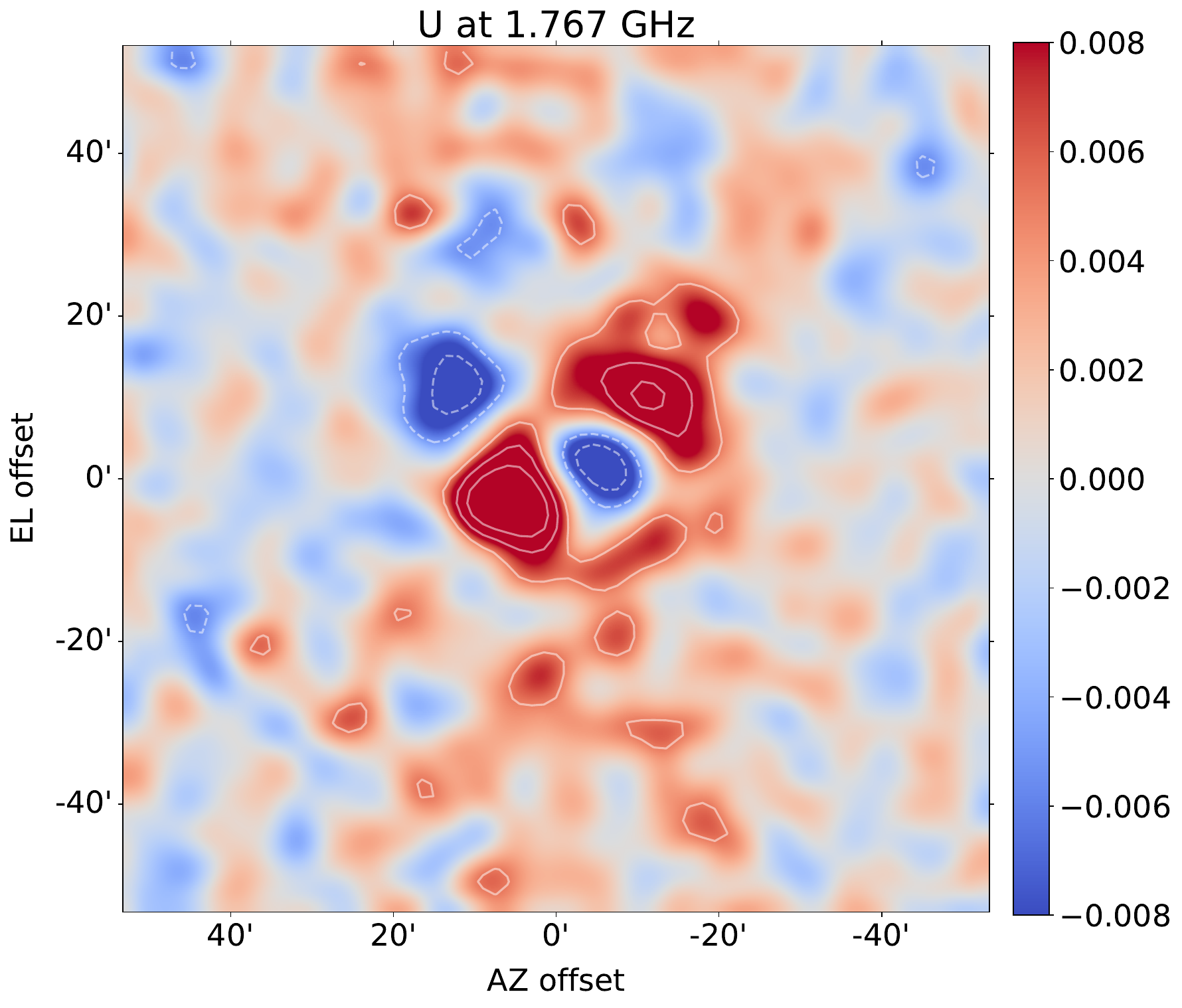}
    \caption{Beam patterns for $I$, $Q$, and $U$ at three frequencies constructed from calibrated scans. The $Q$ and $U$ patterns represent leakage from $I$. All the patterns are normalized to the total intensity peaks. The contour levels for $I$ start at $-3$ dB and stop at $-15$ dB with an interval of 3 dB. For $Q$ and $U$, the contour levels start at $-$0.015 and stop at 0.015 with an interval of 0.005.}
    \label{fig:beam}
\end{figure*}

\begin{table}
\small
\caption{Parameters of STAPS}
\label{tab:param}
\centering
\begin{tabular}{ll}
\hline\hline
Frequency range (observed, MHz) & 1300--1800 \\
Frequency range (usable, MHz)\tablefootmark{a}
                           & 1324--1471, 1482--1492\\
                           & 1506--1524, 1609--1617\\
                           & 1627--1682, 1693--1770\\
Channel width (MHz)              & 1 \\
Number of good channels    & 301 \\
Sky coverage               & $-89\degr<$Dec$<0\degr$\\
Angular resolution (observed)\tablefootmark{b}     &  $19\farcm6-18\farcm1$ (RA)\\
                                        & $16\farcm5-13\farcm8$ (Dec)\\
Angular resolution (smoothed) & $20\arcmin$ \\
Flux density to brightness temperature\tablefootmark{c} & 0.346--0.186 K~Jy$^{-1}$\\
On-axis polarization leakage            &  $<0.8\%$\\
Flux density calibration accuracy       & 10\% \\
RMS sensitivity per channel\tablefootmark{d}  & 16--8~mK for $I$\\
                                              & 8--5~mK for $Q$ and $U$\\
Intensity calibration  &  relative for $I$ \\
                       &  absolute for $Q$ and $U$ \\
\hline
\end{tabular}
\tablefoot{
\tablefoottext{a}{Only frequency gaps larger than 5~MHz are listed.}
\tablefoottext{b}{Based on the beams built from PKS 1934$-$638. The beam slightly varies with declination.}
\tablefoottext{c}{At $20\arcmin$ resolution.}
\tablefoottext{d}{Measured at declination of around $-20\degr$ from the maps smoothed to $20\arcmin$.}}
\end{table}

\section{Observations and data processing}
\label{sec:obs}
The STAPS survey was conducted with Murriyang commensally with the S-PASS survey. During the observations, the S-band ($\lambda$13~cm) receiver was mounted at the primary focus position, while the L-band ($\lambda$20~cm) receiver was placed off-focus. With the same coordinate system as \citet{ruze+65}, where the focus points horizontally along the $z$ axis and the $x$ axis is directed vertically upward, the lateral displacement of the L-band receiver is $x=630$~mm and $z=7.6$~mm. Consequently, the L-band receiver pointed about $1\fdg2$ above the S-band receiver. Data were recorded simultaneously for both surveys. Some of the main parameters of STAPS are listed in Table~\ref{tab:param}. Detailed descriptions of the observation setup and data processing for S-PASS were presented by \citet{carretti+19}. 

The observations are composed of long azimuth (AZ) scans at the fixed elevation (EL) of $33\degr$ for the S-band receiver corresponding to the south celestial pole at Parkes Observatory. All of the observations were conducted at night to avoid the influence of the Sun. Each night, back and forth east scans ($61\fdg5\leq{\rm AZ}\leq 180\degr$) and west scans ($180\degr\leq{\rm AZ}\leq294\degr$) were carried out at a speed of $15\degr$~min$^{-1}$. For STAPS, the fixed elevation was about $1\fdg2$ above the south celestial pole, which delivers a sky coverage of $-89\degr<\rm{Dec}<0\degr$, leaving a gap at the pole. The starting sidereal time for each individual scan was predetermined so that the maximum spacing between two neighboring scans is about $4\farcm4$, slightly less than half of beamwidth at S band to guarantee Nyquist sampling. This means that the STAPS survey is oversampled.

There were eight runs in total, conducted in April, July, October 2008, January, April, July, October 2009, and January 2010, roughly once every 3 months. In total, there are about 7100 east scans and about 3800 west scans after removing the bad scans. The east scans fully sample the sky. The spacing of west scans is twice that of east scans, providing sufficient intersections with east scans to ensure effective destriping. In Fig.~\ref{fig:scans}, some west and east scans in April and October 2008 are illustrated. As can be seen, the combination of east and west scans would produce a large number of crossings~\citep[][their Fig. 3]{carretti+19}, which is required to recover zero levels of intensity in polarization. 

The observation scheme of long AZ scans was developed and demonstrated by \citet{carretti+19} to be able to measure the absolute level of Stokes $Q$ and $U$ and thus the polarized intensity. The long scan results in a large range of parallactic angle, and therefore a modulation of $Q$ and $U$ in the telescope frame. Subtraction of a constant baseline to account for ground emission therefore does not remove $Q$ and $U$ of the sky signal. A joint fitting of all crossings between the west and east scans yields the absolute level of $Q$ and $U$. However, this cannot be applied to the total intensity $I$ since there is no modulation of $I$ against the parallactic angle. Therefore, the zero level is unconstrained in total intensity.  

Data processing followed the procedure developed by \citet[][their Fig.~6]{carretti+19} for the S-PASS survey. All scans were first calibrated and then combined for map-making. PKS B1934$-$638 served as the primary flux calibrator and the flux model by \citet{reynolds+94} was used. PKS B0407$-$658 served as the secondary flux calibrator. Both sources were also used for on-axis leakage calibration, assuming that they are unpolarized. Note that both sources show very small fractional linear polarization of 0.15\% and 0.01\% from the recent observations by \cite{taylor+24}, which will not impact the calibration of polarization leakage. We used PKS 0043$-$424 and 3C~138 to correct the polarization angle. For the former, the intrinsic polarization angle is $137\degr$ with an RM of 2~rad~m$^{-2}$~\citep{broten+88}, and for the latter, the polarization angle is a constant value of $169\degr$~\citep{perley+butler+13}. Each night, both the flux and the polarization calibrators were scanned in the right ascension (RA) and declination (Dec) directions. The gain, on-axis polarization leakage, and instrumental polarization angle were derived, and these calibration solutions were applied to the long AZ scans. 

For map-making, the bad scans that were influenced by strong radio frequency interference (RFI) and thus show large values of root mean square (rms) in total intensity or large values in Stokes $V$ were flagged. Parts of the scans close to the planets were also flagged. Zero-level corrected maps of coarse resolution for the east and west scans were derived by solving intensity levels with crossings~\citep[][Appendix A]{carretti+19}, and a basket-weaving technique by \citet{emerson+graeve+88} was implemented to remove residual discontinuities. The baselines of each calibrated scan were then adjusted using the zero-level corrected map of coarse resolution, and all the adjusted scans were subsequently combined to form the final high resolution maps. The maps cover a declination range from $-89\degr$ to $0\degr$.

The calibration and map-making procedures were performed for each individual frequency channel of 1-MHz width. Of the total of 512 frequency channels, we obtained high-quality maps of $I$, $Q$, and $U$ for 301 channels. Because of low gains, 60 channels at the two ends of the band were flagged. The channels in the frequency range from about 1.52~GHz to about 1.61~GHz were dominated by RFI. For these frequency channels, the data left after flagging were insufficient to make maps. 

\section{Verification}
\label{sec:verification}

\subsection{The beam pattern}
\label{subsec:beam}

 Because of the displacement of the receiver, the beam was slightly distorted with a coma lobe. We collected all calibrated scans close to the unpolarized bright source PKS B1934$-$638 and constructed the beams of $I$, $Q$, and $U$ for each individual frequency channel in the azimuth-elevation frame. The beams at 1.325~GHz, 1.523~GHz, and 1.767~GHz are shown in Fig.~\ref{fig:beam}. Here, all the beam patterns are normalized by the peaks in the total intensity beams. 

For total intensity, the peak of the coma lobe decreases slightly with increasing frequency, ranging from about 9\% at the low frequency to about 6\% at the high frequency. The measured levels of the coma lobe are consistent with electromagnetic simulations of the performance of the telescope with an offset feed~\citep{granet2007}. Note that we obtained the gains based on the fitting of the peaks of the main beams during calibration. For compact sources, the peak intensity is expected to be accurate as shown later, but the estimate of integrated flux density will be influenced by the coma lobe and thus needs a proper deconvolution. For diffuse emission, the intensity level is less influenced by the coma lobe because the emission is smooth. We focus on the large-scale emission from the Galaxy, and therefore did not perform deconvolution for the current data. 

The $Q$ and $U$ patterns in Fig.~\ref{fig:beam} represent the residual leakage from $I$. The centers of the beam patterns are shifted towards the coma lobe. The on-axis residual instrumental polarization is less than about 0.8\% for all frequencies after leakage correction. In contrast with the total intensity, the off-axis residual polarization caused by the coma lobe becomes stronger at higher frequencies, running from about 2\% at 1.325~GHz to about 4\% at 1.767~GHz. The off-axis polarization is mostly canceled out after integration over the beam~\citep{carretti+19}. However, the polarization beam is asymmetric, and instrumental polarization can be generated near strong total intensity emission, such as that of the Galactic plane. We did not correct the off-axis polarization in the current data.  

\begin{figure}
    \centering
    \includegraphics[width=0.48\textwidth]{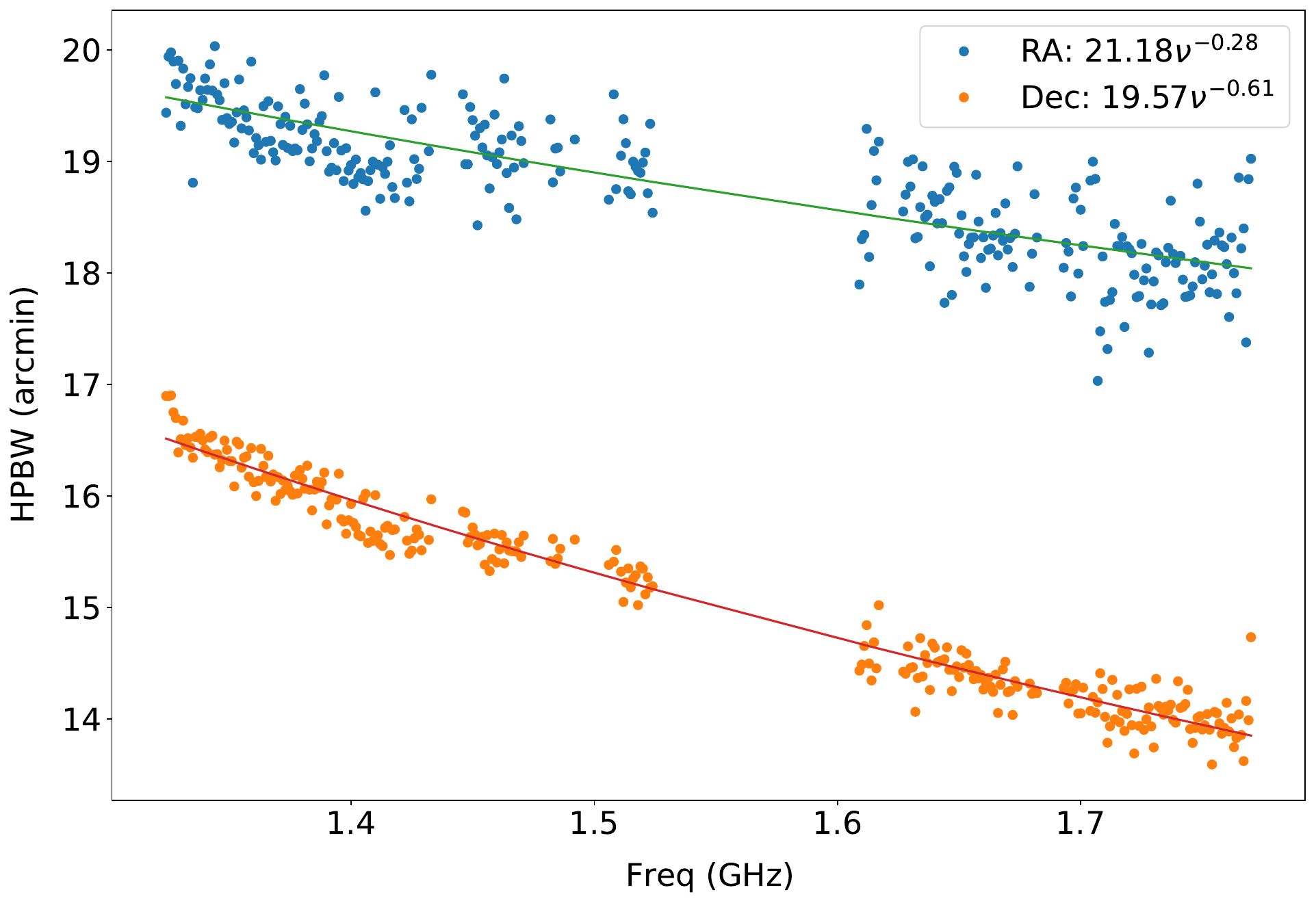}
    \caption{Beamwidth along RA and Dec at different frequencies derived from Gaussian fits to the total intensity beam patterns constructed in the equatorial coordinate.}
    \label{fig:beam_width}
\end{figure}

\begin{figure}
    \centering
    \includegraphics[width=0.48\textwidth]{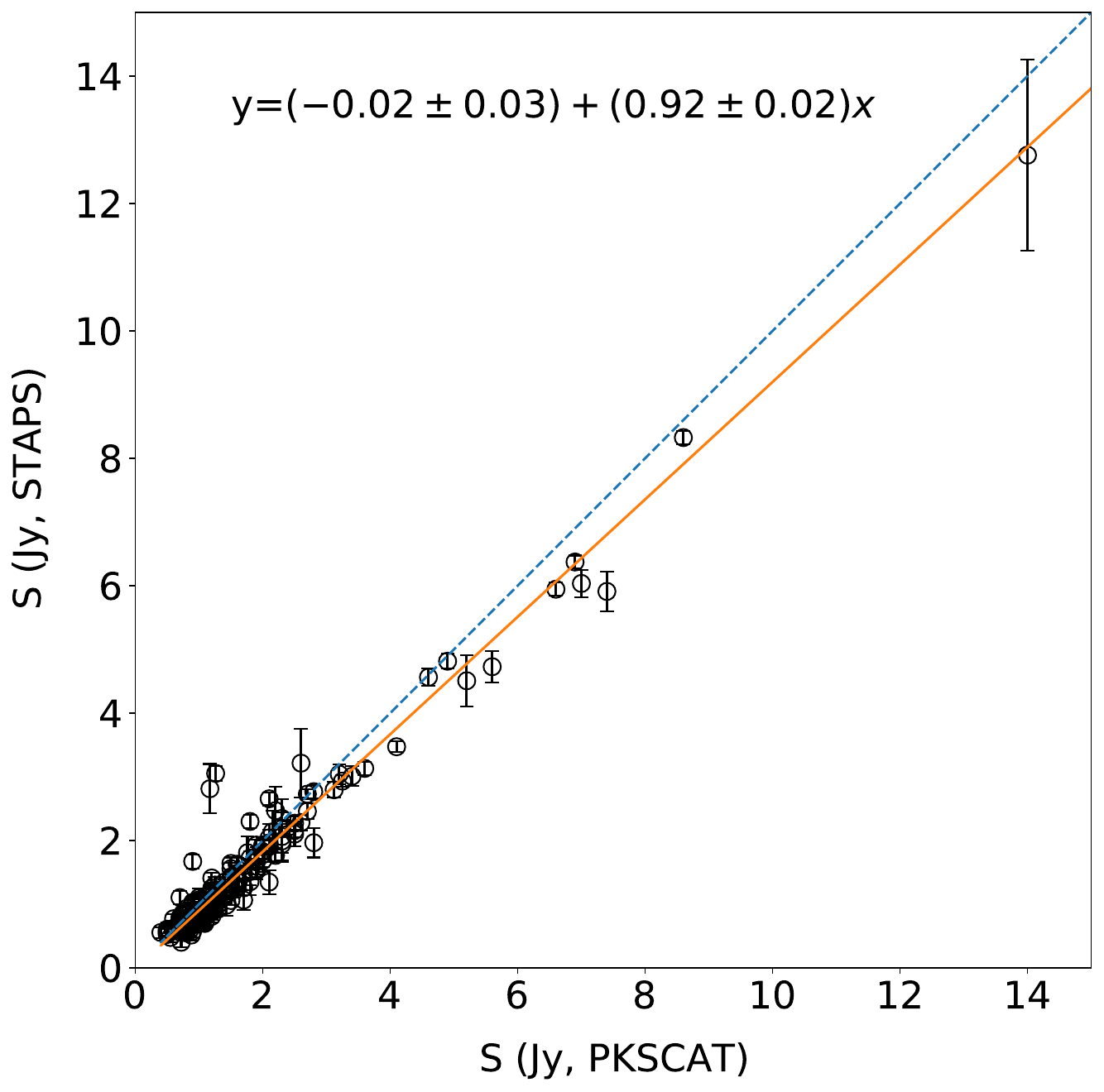}
    \caption{Comparison of flux densities between STAPS and PKSCAT. The dashed line marks $y=x$ and the solid line is a linear best fit with the parameters showing in the figure.}
    \label{fig:source_flux}
\end{figure}

The beam is non-circular in the azimuth-elevation frame~(Fig.~\ref{fig:beam}). When the telescope is moving along AZ direction for an east scan, the beam is not aligned north-south in the equatorial frame. Instead, it is inclined to that frame at an angle, ${\beta}_E$, which can be determined from the parallactic angle, and can be as large as ${\pi}/4$. When the telescope is moving along AZ direction for a west scan, the inclination of the beam, ${\beta}_W$, in the equatorial frame is very different, sometimes even orthogonal to ${\beta}_E$. When the data from the east and west scans are combined in the map-making process in the equatorial system, the beam will vary across the sky because of the changing difference between ${\beta}_E$ and ${\beta}_W$. Near the south pole, east and west scans are close to parallel, so the beam in the map will be roughly aligned north-south, but far from the south pole the form of the beam in the maps becomes harder to predict. 

We have approached this problem empirically. We constructed the total intensity beam patterns from PKS 1934$-$638, which are similar to those shown in Fig.~\ref{fig:beam} but in the equatorial frame. We then obtained the half power beamwidth (HPBW) by fitting 2D Gaussian functions to these total intensity beam patterns. The result is shown in Fig.~\ref{fig:beam_width}. The beam in right ascension is always larger than the beam in declination. The beamwidth versus frequency can be fitted to a power law, as indicated in Fig.~\ref{fig:beam_width}. We also examined the beam patterns constructed from other sources at different declinations and found that the beam is surprisingly uniform. The measured beamwidths in right ascension show a greater scatter; scatter in declination is about 2\% and in right ascension 10\%. We have adopted the beam dimensions shown by the fitted lines in Figure 3 and smoothed the maps to a resolution of $20\arcmin$ at all frequencies. The variation of the beam versus declination introduced an error of less than 10\% for the flux density scale. Smoothing, as described above, made only small changes in the appearance of the maps, indicating that most of the diffuse structure has scales larger than the beam.

\subsection{The flux density scale}
The observation and data processing procedures were optimized for large-scale emission, but it is still worthwhile to examine the flux densities of compact sources. We performed source finding with {\scriptsize AEGEAN}~\citep{hancock+18P} on the STAPS total intensity image at 1.41~GHz, and cross matched the results with the Parkes Southern Radio Source Catalog~\citep[PKSCAT,][]{wright+90}. The comparison of the flux densities is shown in Fig.~\ref{fig:source_flux}. Note that the peak flux densities from STAPS were used here to avoid the influence of coma lobes.

As can be seen in Fig.~\ref{fig:source_flux}, the flux densities of STAPS and PKSCAT agree reasonably well. A linear fitting yields a flux density ratio of 0.92, close to 1. This means an uncertainty of less than about 10\% for the flux densities of compact sources from STAPS.  

To examine the flux density scales of both total intensity and polarized intensity for large-scale structures, we turned to radio galaxy Cen~A, which is an extended and polarized source. We obtained the images of Cen~A observed with Parkes by \citet{o'sullivan+13}, and smoothed them to a resolution of $20\arcmin$. We made pixel intensity versus pixel intensity plots at three frequencies for $I$, $Q$, $U$ and polarized intensity (Fig.~\ref{fig:cena}) toward the southern lobe of Cen~A~\citep[][their Fig.~1]{o'sullivan+13}.  

As can be seen in Fig.~\ref{fig:cena}, there is good agreement between STAPS and the Parkes observations for total intensity and polarization. The linear fits show that the pixel intensity ratio is around 1 with an uncertainty of about 10\%. The comparison also confirms that the coma lobes have little influence on large-scale emission. Note that there are offsets in the fits, which are due to the different zero levels. 

We conclude that the flux density scale of the STAPS data is accurate within 10\%. 

\subsection{The main beam brightness temperature scale}
\label{subsec:scale} 

We do not have the aperture efficiency and main beam efficiency measured for the L-band receiver at the off-focus position, and thus could not establish a temperature standard solely from STAPS. Fortunately, the GMIMS high-band north survey, covering virtually the same frequency range and overlapping with STAPS for a large area with a declination range between $-30\degr$ and $0\degr$, was absolutely calibrated to main beam brightness temperature~\citep{du+2016,wolleben+21}. This allowed us to derive the temperature scale by comparing STAPS with the GMIMS high-band north.

We selected the Galactic plane area $0\degr\lesssim l \lesssim 30\degr$ and $|b|\lesssim5\degr$ where the total intensity emission is bright for both GMIMS and STAPS. We smoothed the STAPS data to the GMIMS resolution of $40\arcmin$, made pixel intensity versus pixel intensity plots for total intensity $I$ at each individual frequency, and performed linear fits to obtain the conversion factors from flux density in Jy beam$^{-1}$ to main beam brightness temperature in K. The results for randomly selected frequencies between 1.3 and 1.8~GHz are shown in Fig.~\ref{fig:scales}. The data points are very tight around a linear relation, which verified the map-making procedure of STAPS. The conversion factors at $20\arcmin$ resolution were then derived and are in the range of 0.346--0.186~K~Jy$^{-1}$ for the frequency range of 1.324--1.770~GHz~(Table~\ref{tab:param}).

\begin{figure}
    \centering
    \includegraphics[width=0.48\textwidth]{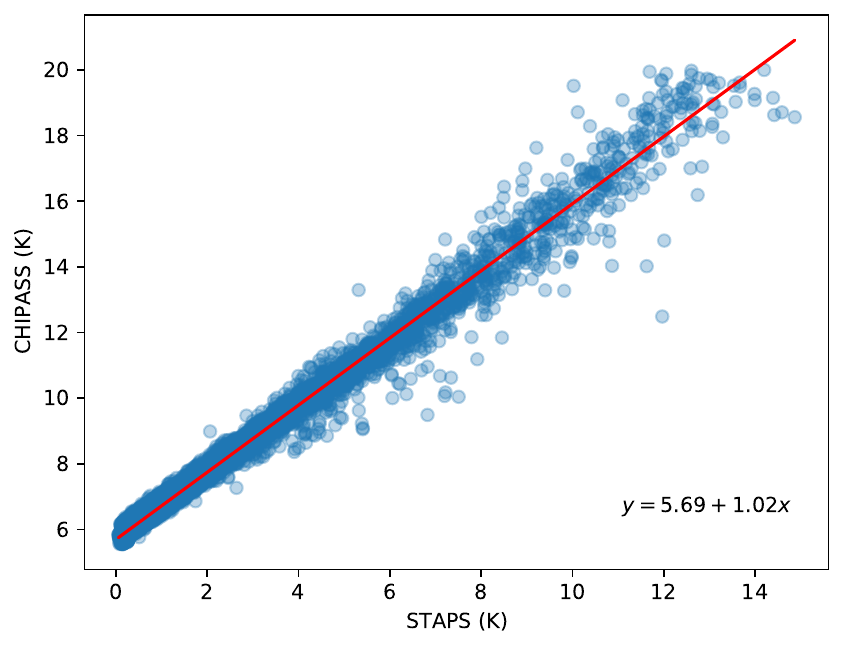}
    \caption{Temperature -- temperature plots between STAPS and CHIPASS at 1.4~GHz. }
    \label{fig:staps_chipass}
\end{figure}

We applied the conversion factors to the STAPS data to obtain the temperature scale. To verify the temperature scale, we retrieve the Continuum HI Parkes All-Sky Survey~\citep[CHIPASS,][]{calabretta+14} image at 1.4~GHz and smoothed it to the STAPS resolution of $20\arcmin$. Note that CHIPASS is in the full beam brightness temperature scale, tied to the 1420~MHz all-sky survey~\citep{reich+82,reich+86,reich+01}. A factor of 1.55~\citep{reich+88} was applied for conversion to main beam brightness temperature. We then made the temperature-temperature plot for the sky area $-89\degr<{\rm Dec}<0\degr$~(Fig.~\ref{fig:staps_chipass}). There is also a tight linear relation. The linear fitting yields a ratio of about 1.02, meaning that the temperature scale of STAPS derived on the basis of GMIMS is reliable.   

\section{Maps}
\label{sec:maps}

\subsection{Total intensity and polarization maps}
\begin{figure}
    \centering
    \includegraphics[width=0.48\textwidth]{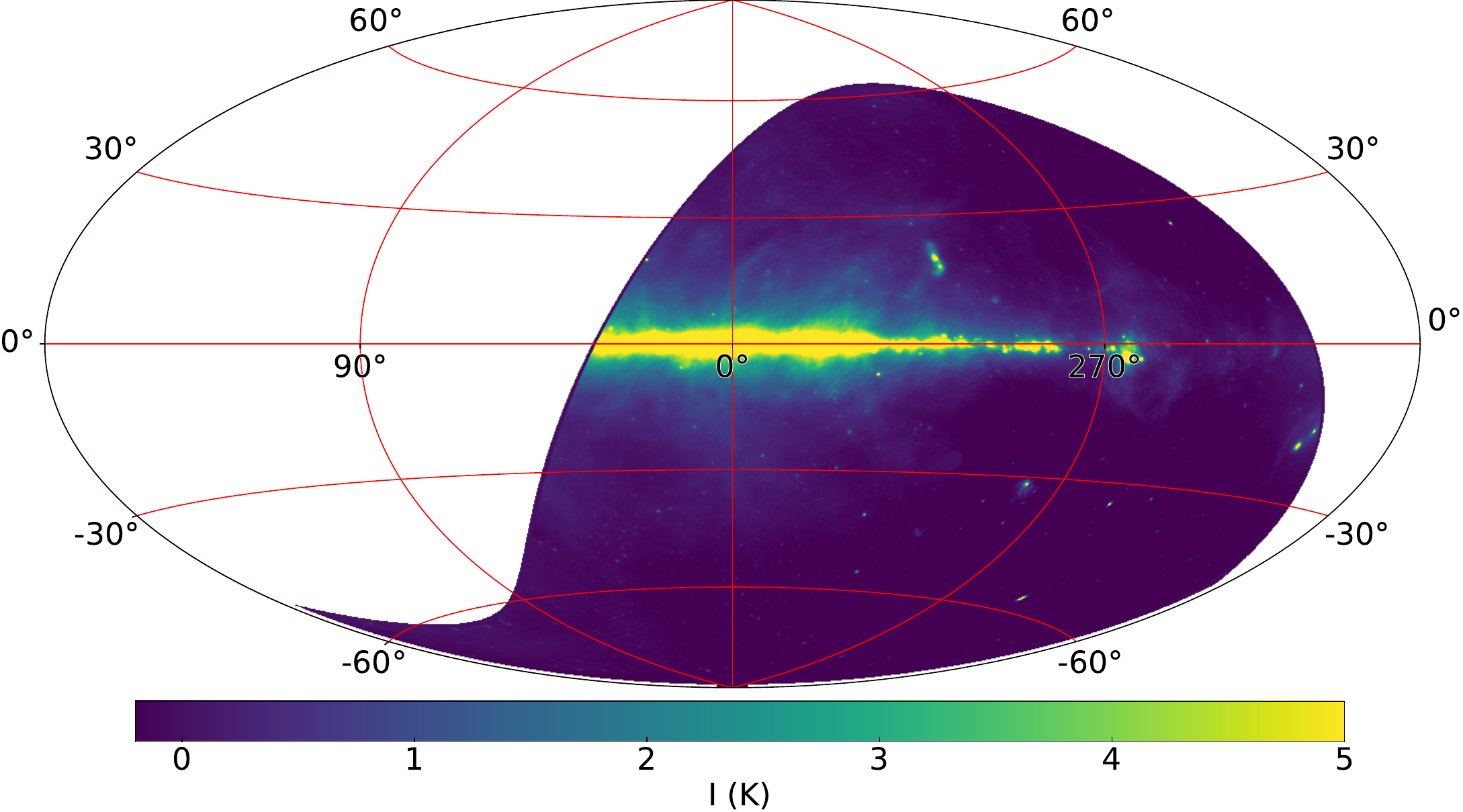}\\[3mm]
    \includegraphics[width=0.48\textwidth]{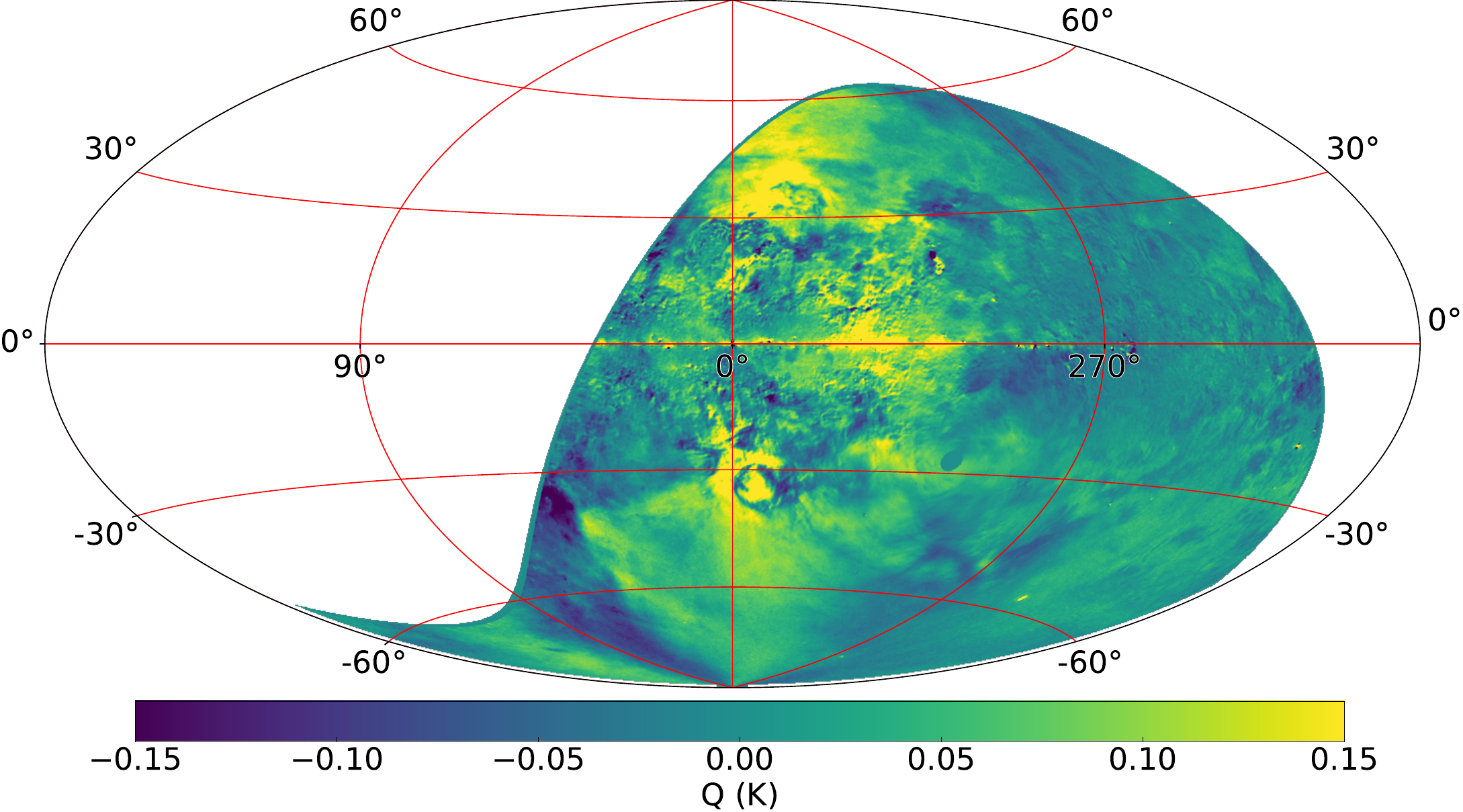}\\[3mm]
    \includegraphics[width=0.48\textwidth]{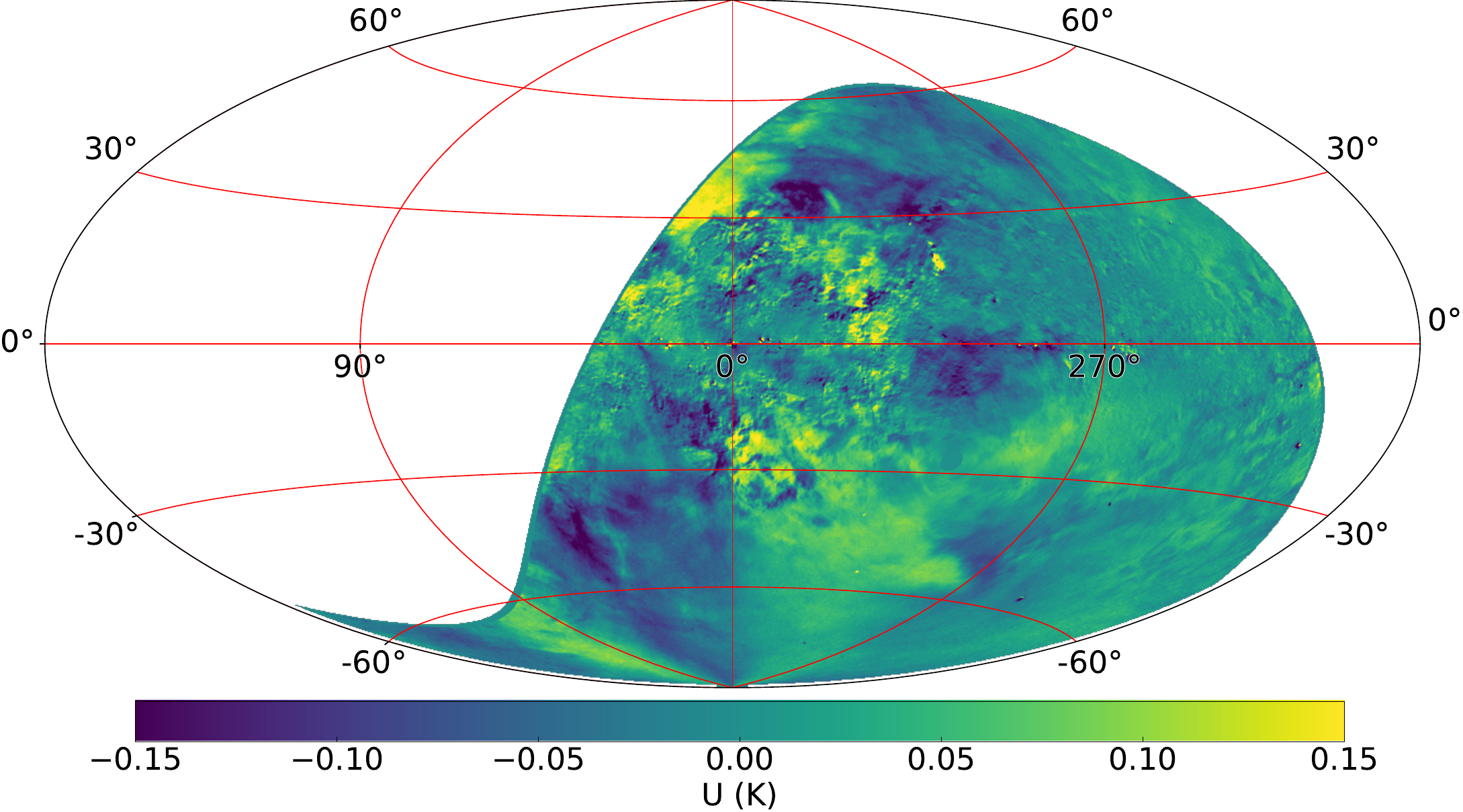}
    \caption{$I$, $Q$, and $U$ maps at 1.45~GHz from STAPS. The resolution is $20\arcmin$.}
    \label{fig:iqu}
\end{figure}

We obtained $I$, $Q$, and $U$ maps at 301 frequency channels in both Jy beam$^{-1}$ and K. All maps have been smoothed to a common resolution of $20\arcmin$. As an example, the maps at 1.45~GHz are shown in Fig.~\ref{fig:iqu}. The total intensity is very smooth with strong emission at the Galactic plane. In contrast, there are many structures of various angular scales in the $Q$ and $U$ images. Because of depolarization, there is virtually no polarized emission expected near the Galactic plane. Polarized emission near some parts of the plane is probably caused by residual instrumental leakage.  

\begin{figure}
    \centering
    \includegraphics[width=0.48\textwidth]{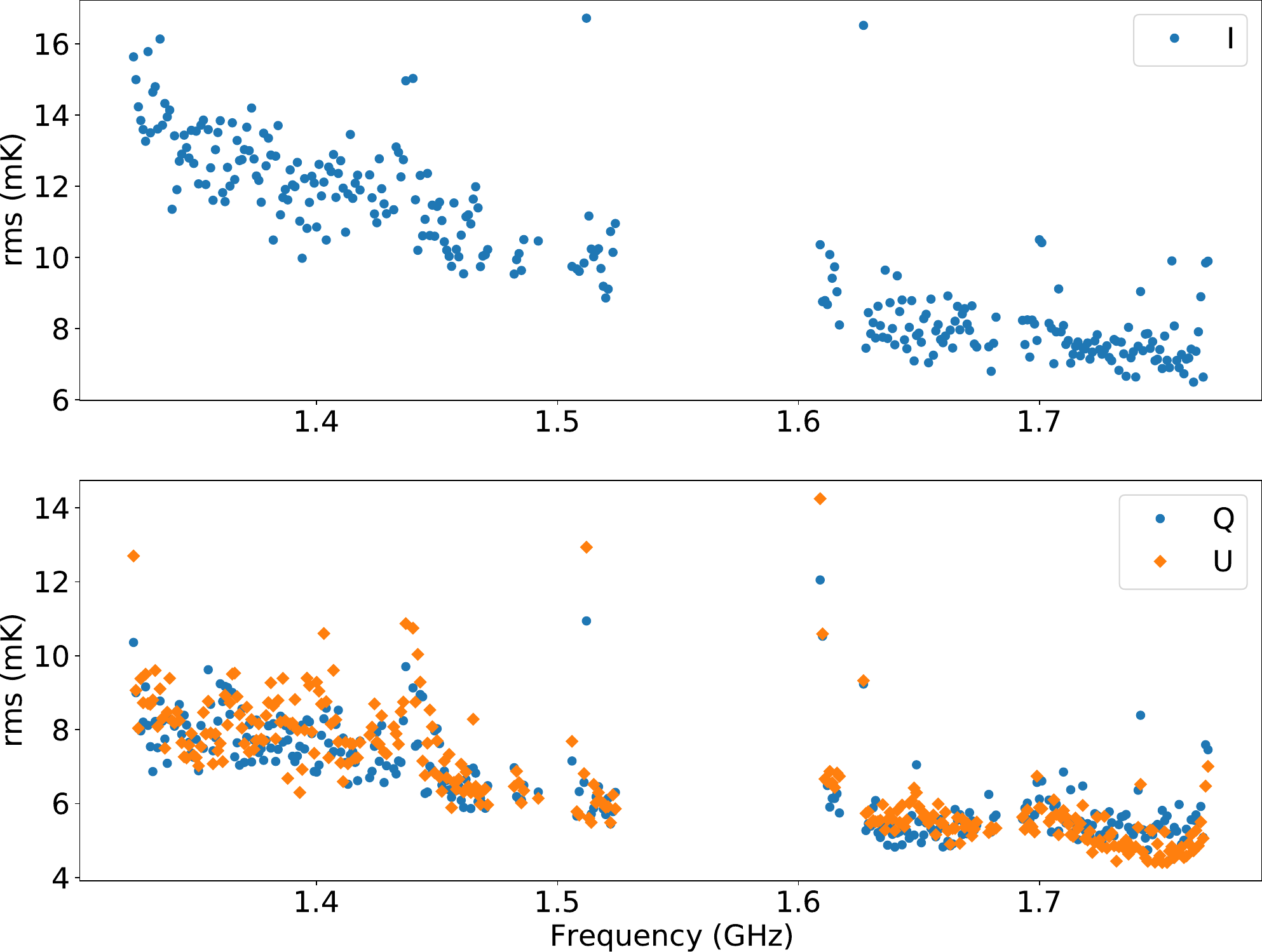}
    \caption{Root mean square (rms) noise levels per channel for $I$, $Q$, and $U$ measured from an area without strong emission at declination of around $-20\degr$.}
    \label{fig:rms}
\end{figure}

We selected areas at declination of around $-20\degr$ where there are no emission structures and calculated the root mean square (rms) noise level in $I$, $Q$ and $U$ for each individual 1-MHz channel. The results are shown in Fig.~\ref{fig:rms}. For $I$, the rms is about 16~mK at the low frequency end and about 8~mK at the high frequency end. For $Q$ and $U$ the rms decreases from about 8~mK to about 5~mK with increasing frequency. 

The number of crossings varies with the declination with more data points within a beam toward the lower declination, as can be seen in Fig.~\ref{fig:scans}. This means that the rms noise depends on declination and reaches the minimum at the declination of $-89\degr$. 

However, we estimated the rms noise in $I$ from different declinations and found no large variations. According to \citet{condon+74} and \citet{meyers+17}, the confusion limit is about 65.7~mJy~beam$^{-1}$ at 1.324~GHz and about 53.6~mJy~beam$^{-1}$ at 1.770~GHz at a resolution of $20\arcmin$, corresponding to about 23~mK and 10~mK, respectively. These values are slightly higher than what we measured, indicating that the total intensity probably has already hit the confusion limit, and thus the rms noise is independent of declination. 

\begin{figure}
    \centering
    \includegraphics[width=0.48\textwidth]{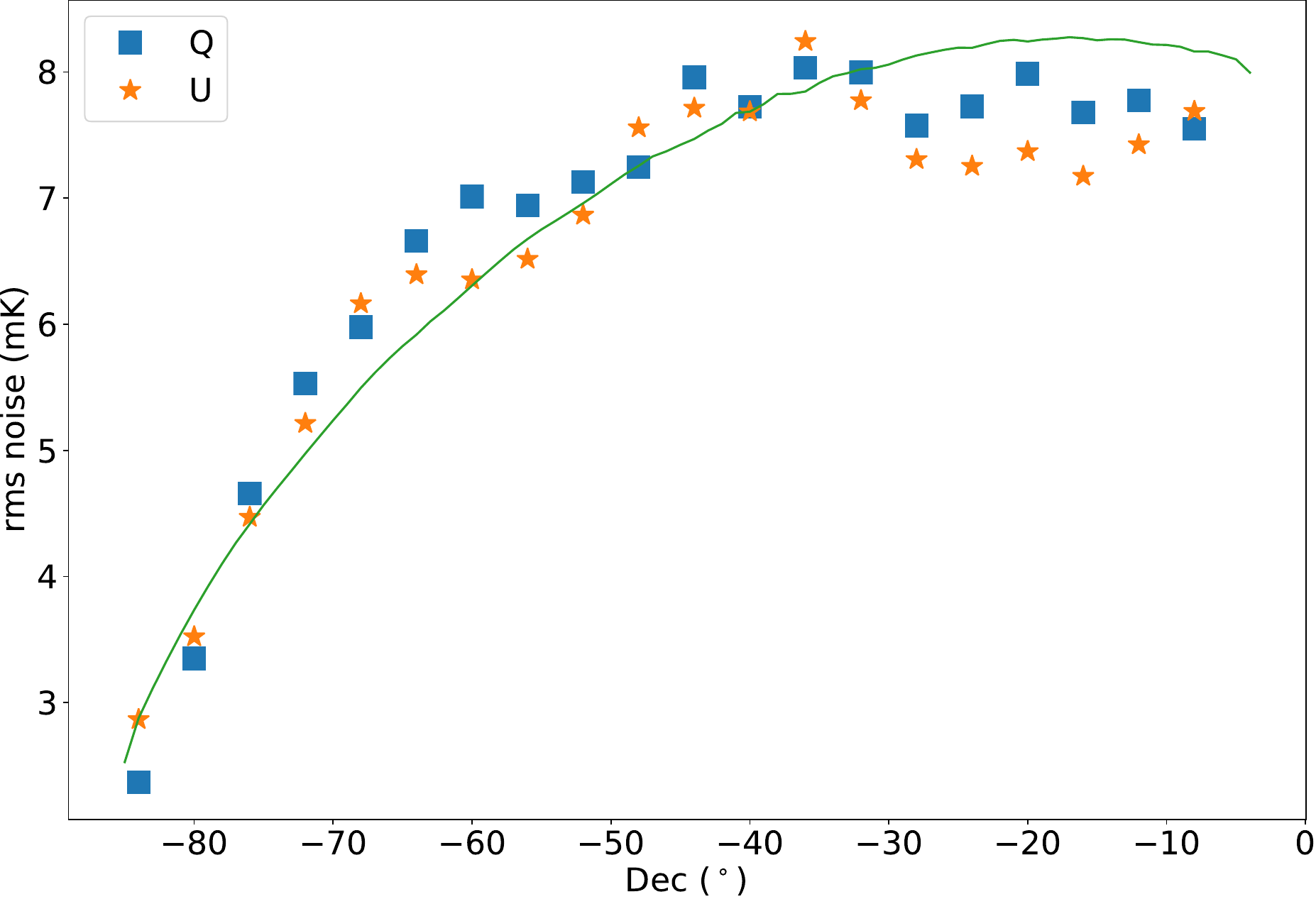}
    \caption{Root mean square (rms) noise per channel in $Q$ and $U$ versus declination. The line is derived from the number of crossing points within a beam.}
    \label{fig:rms_dec}
\end{figure}

For $Q$ and $U$, the emission structures from the Galactic interstellar medium are distributed almost everywhere, which makes it difficult to examine the dependence of rms noise on declination. We obtained the difference images of two $Q$ or $U$ images with consecutive frequencies. In this way, the Galactic emission that is common in both images is removed, and only the noise remains. We then obtained the rms noise of $Q$ and $U$ versus the declination shown in Fig.~\ref{fig:rms_dec}. We also obtained the number of crossing points ($N$) within a beam from map-making. The theoretical rms is expected to be proportional to $1/\sqrt{N}$, and is also plotted in Fig.~\ref{fig:rms_dec}. It can be clearly seen that the measured rms noise in $Q$ and $U$ is consistent with the theoretical expectation.

\subsection{RM synthesis maps}
The observed polarization as a function of frequency $\nu$ or wavelength squared $\lambda^2$ can be written in a complex quantity as $P(\lambda^2)=Q(\lambda^2)+iU(\lambda^2)$. Polarized emission from a source can be described with a Faraday depth spectrum as $F(\varphi)=Q(\varphi)+iU(\varphi)$, where $\varphi$ is the Faraday depth and $|F(\varphi)|$ gives the polarized intensity residing at $\varphi$. These two quantities $P(\lambda^2)$ and $F(\varphi)$ are Fourier transform pairs, and RM synthesis, developed by \citet{Burn66,brentjens+05}, can be used to derive $F(\varphi)$ from $P(\lambda^2)$. Because of the limited sampling in the $\lambda^2$ domain, the observed $F(\varphi)$ is a convolution of the true $F(\varphi)$ with a window function. The deconvolution can be performed with the RM clean method developed by \citet{heald+09}.

\begin{figure}
    \centering
    \includegraphics[width=0.48\textwidth]{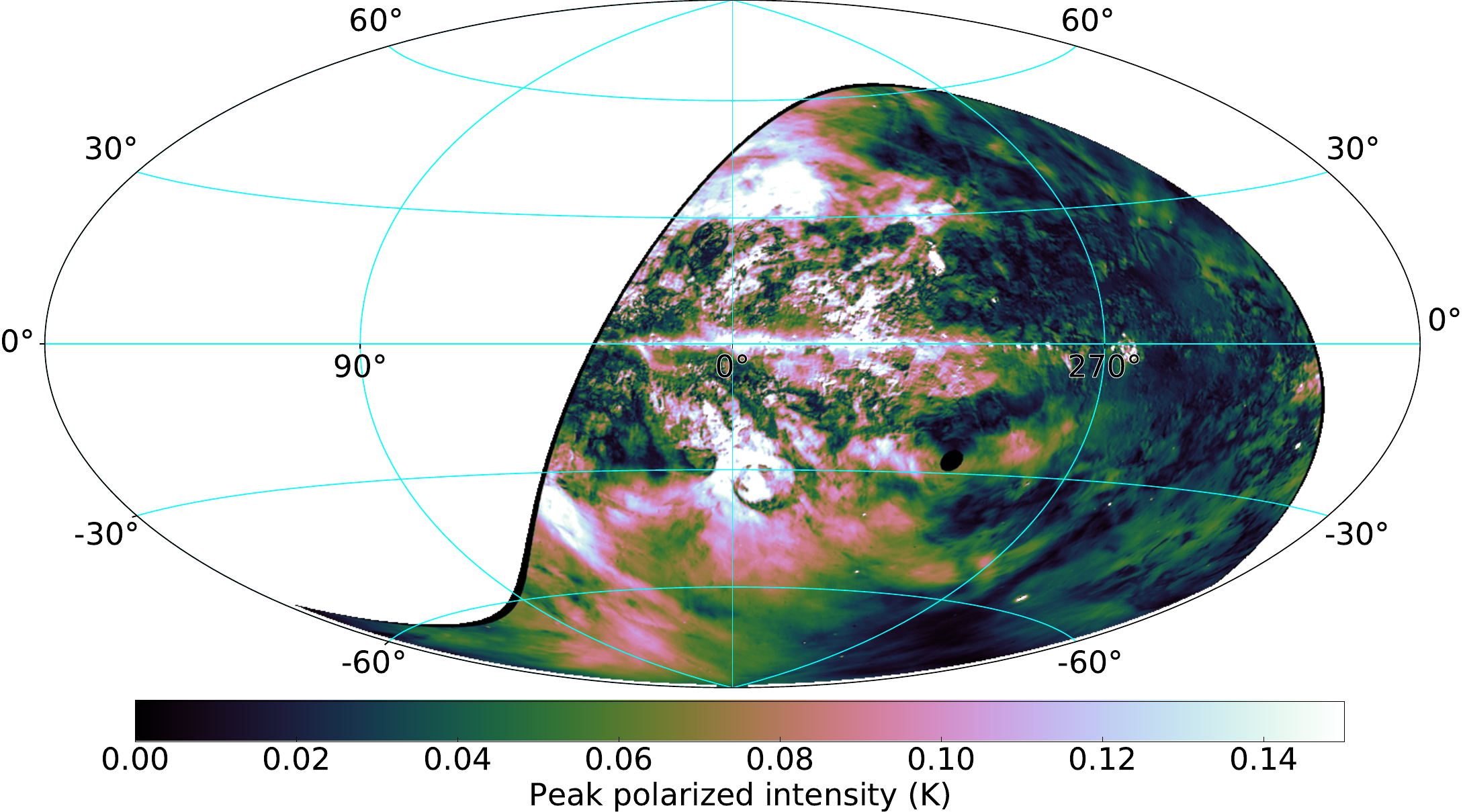}\\[3mm]
    \includegraphics[width=0.48\textwidth]{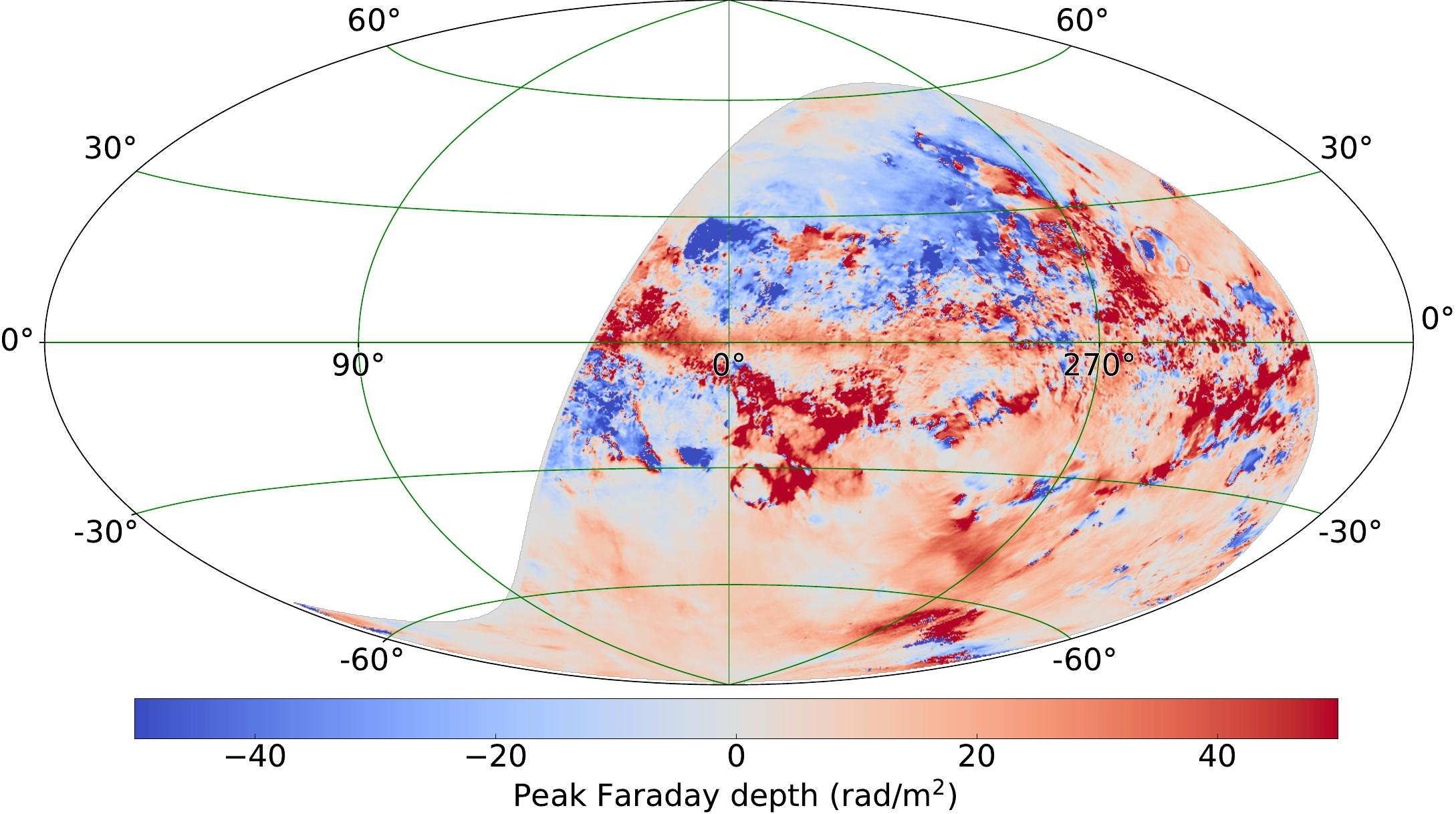}
    \caption{Peak polarized intensity and Faraday depth derived from Faraday spectra.}
    \label{fig:prm_rms}
\end{figure}

We apply RM synthesis~\citep{brentjens+05} and RM clean~\citep{heald+09} to obtain the Faraday spectra. We then search for the peak of $|F(\varphi)|$ for each individual pixel. The polarized intensity and Faraday depth of peaks are shown in Fig.~\ref{fig:prm_rms}. The Faraday depth of the peak can be regarded as the RM if the Faraday spectrum is simple with only one peak. This is the case for most of the pixels, but see \citet{raycheva+24} for a discussion on complex Faraday spectra. From the image of peak polarized intensity, we can clearly see a boundary at Galactic latitude $|b|\approx30\degr$ below which extended, bright polarized emission at higher absolute latitudes is abruptly depolarized. Note that there is still polarized emission towards the Galactic plane, which is caused by instrumental leakage.

\citet{schnizerler+2019} identified polarized sources from S-PASS and observed these sources with Australian Telescope Compact Array at 1.1-3.1~GHz to determine their RMs with the QU-fitting method~\citep[e.g.][]{sun+15b}. We extracted Faraday spectra towards the sources observed by \citet{schnizerler+2019}, and derived the RMs for sources with peak polarized intensity larger than the local background by three times the local rms. Here, we used only Faraday simple spectra that have a single dominant peak. 

\begin{figure}
    \centering
    \includegraphics[width=0.48\textwidth]{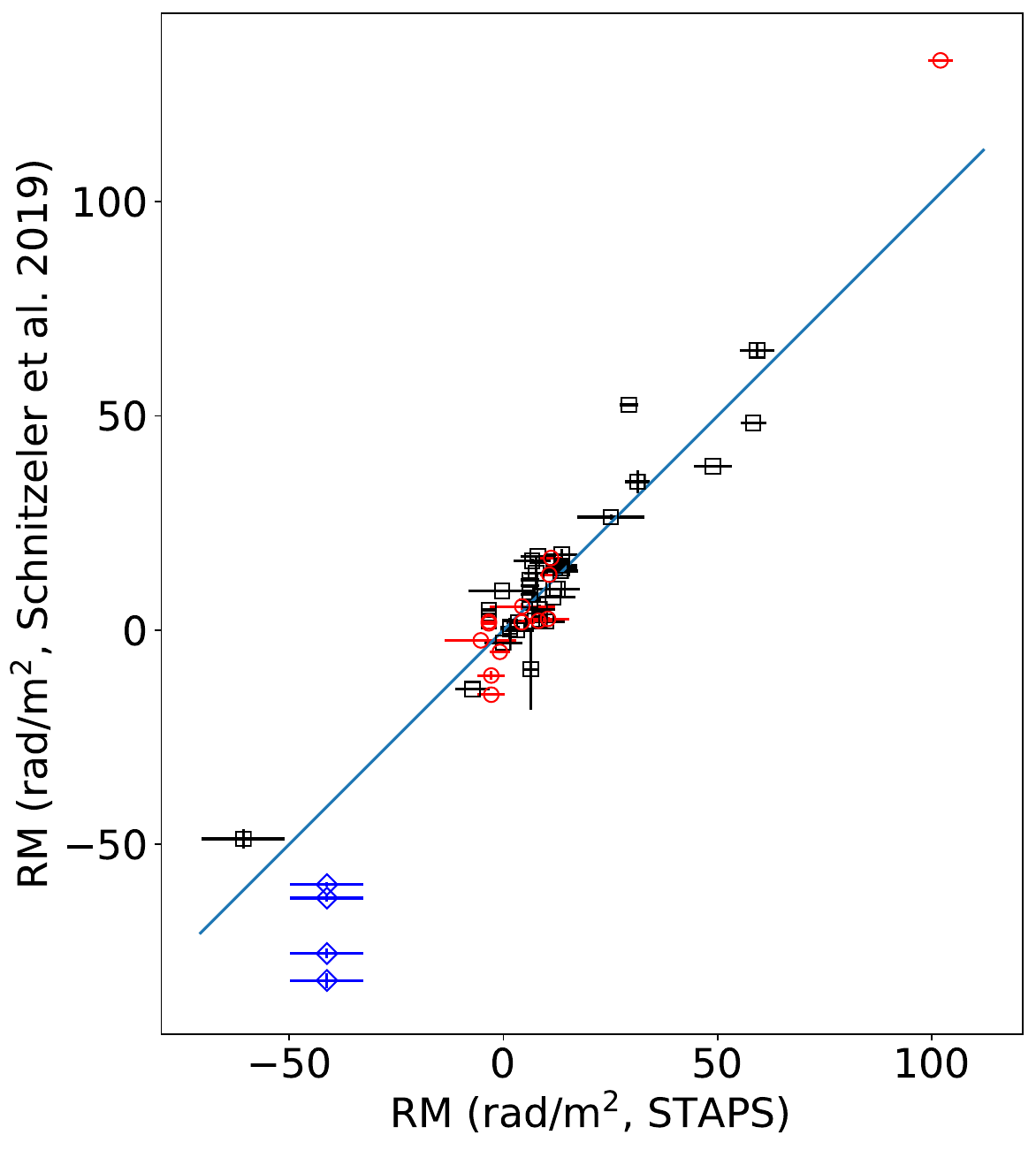}
    \caption{Comparison of RMs derived from STAPS and the measurements by \citet{schnizerler+2019}. The sources with only one RM component in \citet{schnizerler+2019} are marked with red circles. For sources with multiple RM components, RMs of the first components are plotted and marked with black squares, except for one source with multiple observations, RMs of the second components are plotted and marked with blue diamonds. The line indicates $y=x$.}
    \label{fig:rm_compare}
\end{figure}

The comparison of the RMs is shown in Fig.~\ref{fig:rm_compare}. Based on the results by \citet{schnizerler+2019}, many sources have more than one RM component. In this case, we used RMs of the first components that have the highest fractional polarization; except for one source with multiple observations, RMs of the second component were closer to the RM from STAPS and were used. As we can see, the RMs generally agree with each other. This implies that RM from both the single component of simple sources and the first component of complex sources is mainly contributed by the Galactic interstellar medium other than that from within the sources. For the source with multiple observations, the RM varies between different observations, and the variation is much larger than the errors~(Fig.~\ref{fig:rm_compare}), indicating a variable nature of the source in the time domain that needs further investigation.  

\section{Conclusions}
\label{sec:con}

STAPS is the first broadband multichannel polarization survey of the southern sky ($-89\degr<{\rm Dec}<0\degr$) at L-band covering the frequency range of 1.3--1.8 GHz. STAPS was observed commensally with the S-PASS survey. The unique feature for S-PASS and STAPS is that the surveys are composed of long azimuth scans, enabling us to achieve zero-level calibration for $Q$ and $U$. The data processing procedure developed by \citet{carretti+19} for S-PASS was used to process STAPS. We obtained high-quality $I$, $Q$, and $U$ maps for 301 frequency channels each with 1~MHz width, smoothed to a common resolution of $20\arcmin$.

The consistency of the flux densities of compact sources between STAPS and PKSCAT, and the good correspondence of the pixel values for total intensity and polarized intensity of Cen~A with a ratio close to 1 between STAPS and the Parkes observations by~\citet{o'sullivan+13} indicate that the flux density scale of STAPS is reliable and the influence of the coma lobes on both total intensity and polarized intensity of extended emission is small. 

We derived the conversion factors from flux density (Jy beam$^{-1}$) to main beam brightness temperature (K) using the GMIMS high-band north survey and applied these factors to STAPS data. The temperature-temperature plot between STAPS and CHIPASS exhibits a tight linear relation with a gradient close to 1, indicating a reliable temperature scale and map-making procedure for STAPS.

We also ran RM synthesis and RM clean to obtain Faraday depth cubes. From the peak Faraday depth or RM map, we obtained RMs of extragalactic sources whose RMs were derived by \citet{schnizerler+2019} from ATCA observations at 1.1--3.1~GHz. The agreement of these RMs indicates that the polarization calibration and map-making for STAPS are also reliable.

The STAPS cubes together with those of the GMIMS high-band north survey provide a multifrequency polarization view of the entire Milky Way Galaxy, which will help us understand the Galactic magnetic field and magnetized interstellar medium.

\section{Data availability}
The image cubes for Stokes $I$, $Q$, and $U$ in Equatorial coordinates with a common resolution of $20\arcmin$ are available in electronic form at the CDS via anonymous ftp to cdsarc.u-strasbg.fr (130.79.128.5) or via http://cdsweb.u-strasbg.fr/cgi-bin/qcat?J/A+A/. These data can also be accessed from the Canadian Advanced Network for Astronomical Research (doi: 10.11570/25.0004) via https://doi.org/10.11570/25.0004.

\begin{acknowledgements}
We thank the referee for the comments that have improved the paper. We thank Drs. Paddy Leahy, Ann Mao, and Anna Ordog, and Prof. John Dickey for valuable comments on the paper. XS is supported by the National Natural Science Foundation of China (No. 12433006) and the National SKA Program of China (Grant No. 2022SKA0120101). This work is part of the joint NWO-CAS research program in the field of radio astronomy with project number 629.001.022, which is (partly) financed by the Dutch Research Council (NWO). MH acknowledges funding from the European Research Council (ERC) under the European Union's Horizon 2020 research and innovation program (grant agreement No 772663). Murriyang, the Parkes 64m radio-telescope is part of the Australia Telescope National Facility which is funded by the Australian Government for operation as a National Facility managed by CSIRO. We acknowledge the Wiradjuri people as the Traditional Owners of the Observatory site.
\end{acknowledgements}

\bibliographystyle{aa}
\bibliography{bibtex} 
\begin{appendix}
\onecolumn
\section{Flux density comparison}
\begin{figure*}[h!]
    \centering
    \includegraphics[width=0.235\textwidth]{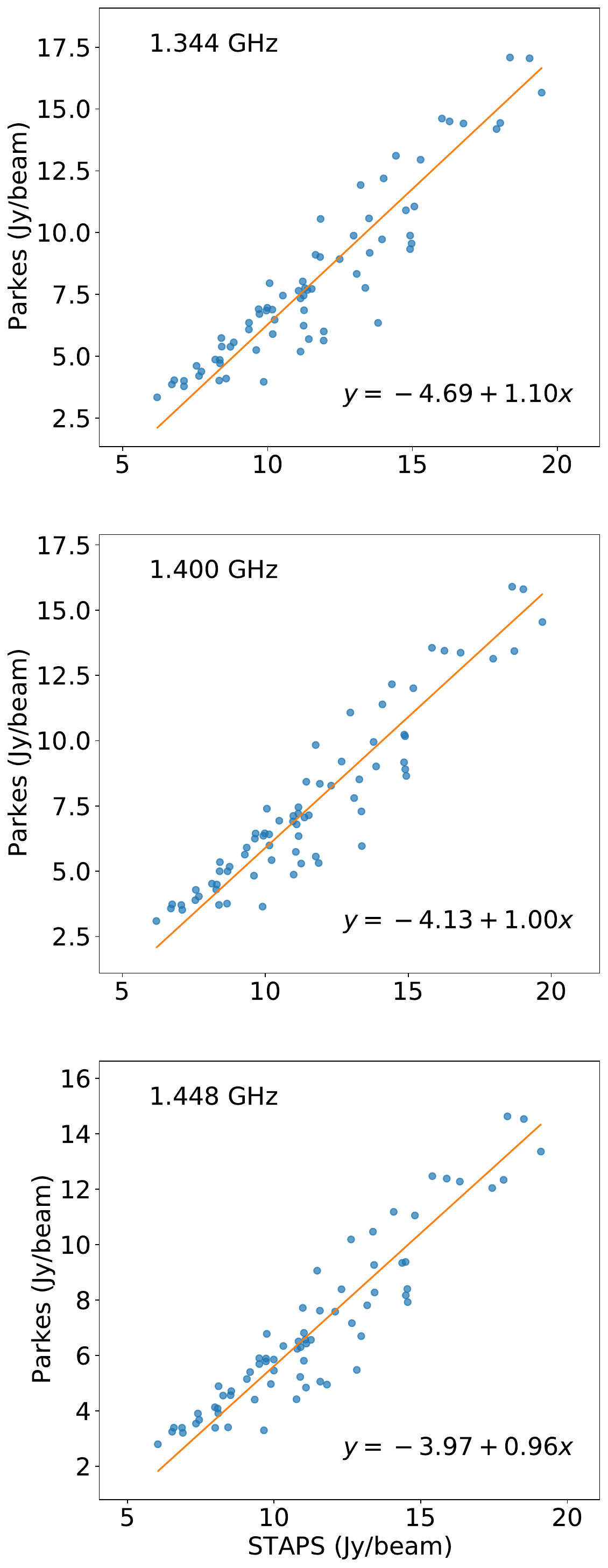}
    \includegraphics[width=0.235\textwidth]{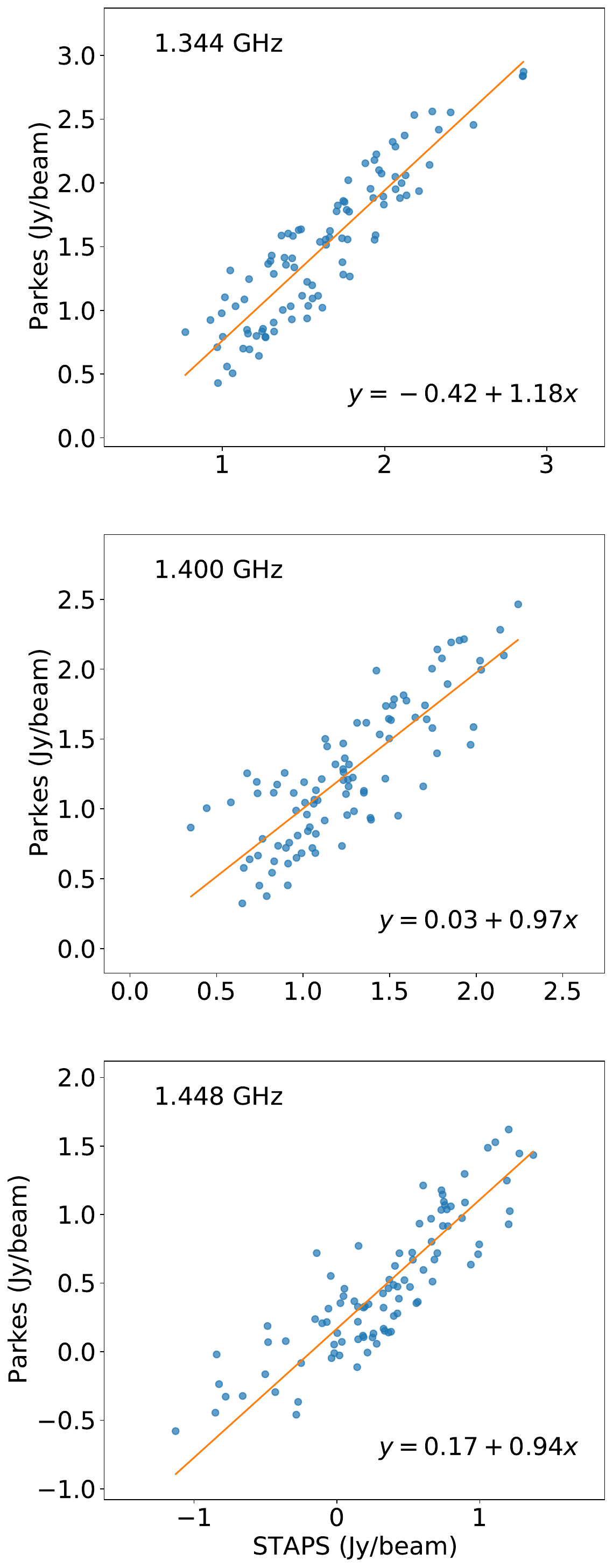}
    \includegraphics[width=0.23\textwidth]{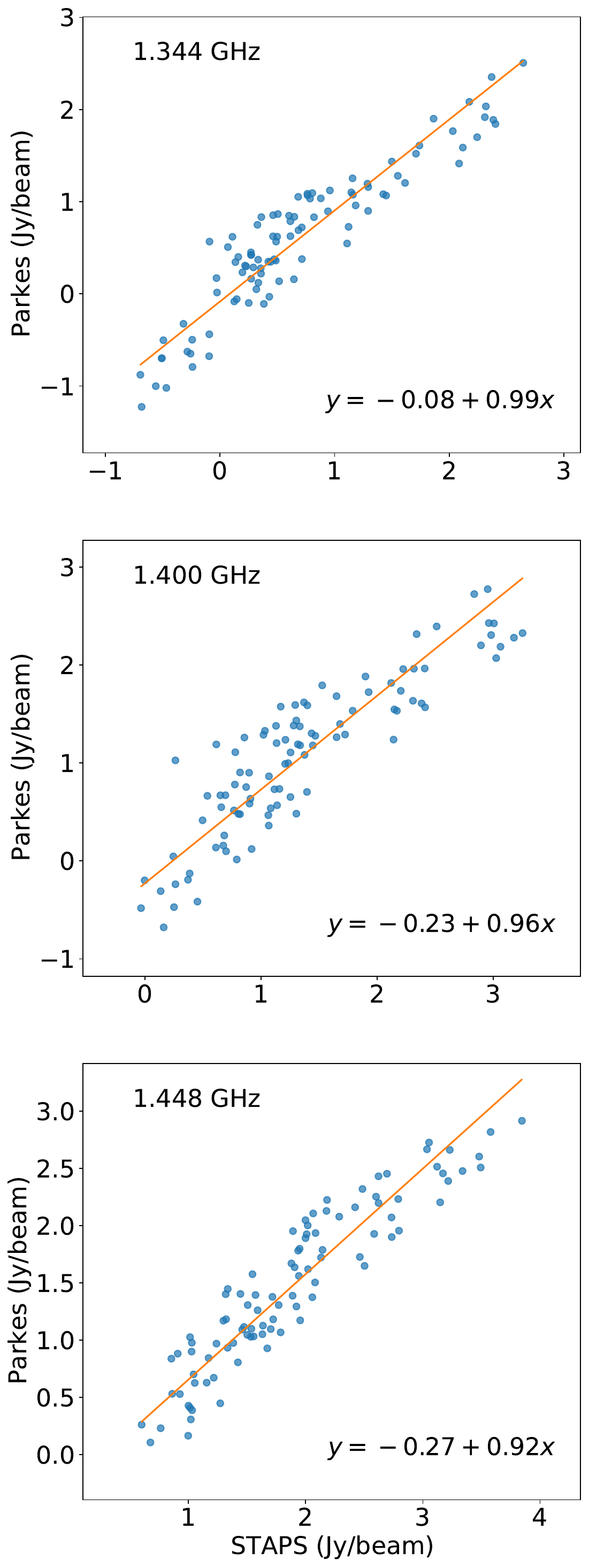}
    \includegraphics[width=0.23\textwidth]{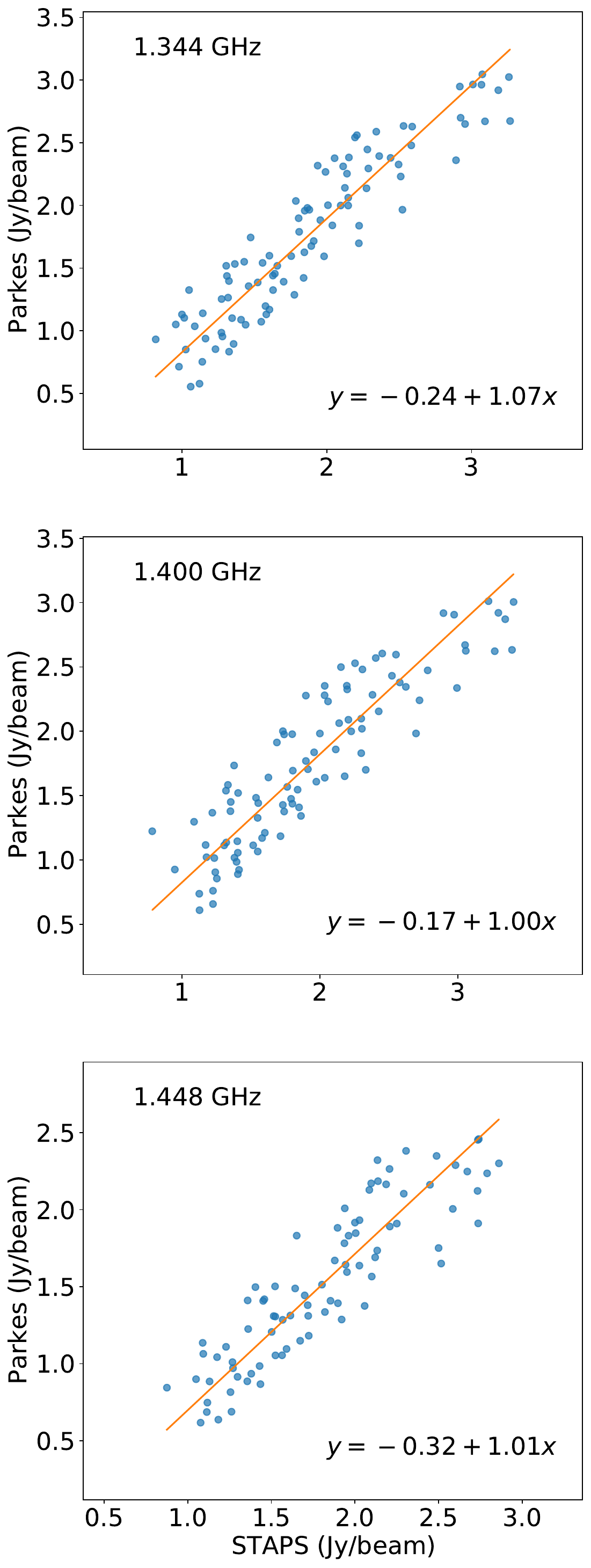}
    \caption{Comparison of pixel densities of Cen~A between STAPS and the Parkes observation by \citet{o'sullivan+13} at three frequencies (from top to bottom). From left to right are: $I$, $Q$, $U$, and polarized intensity. The solid lines indicate linear fits.}
    \label{fig:cena}
\end{figure*}

\onecolumn
\section{Brightness temperature versus flux density}

\begin{figure*}[h!]
    \centering
    \includegraphics[width=0.9\textwidth]{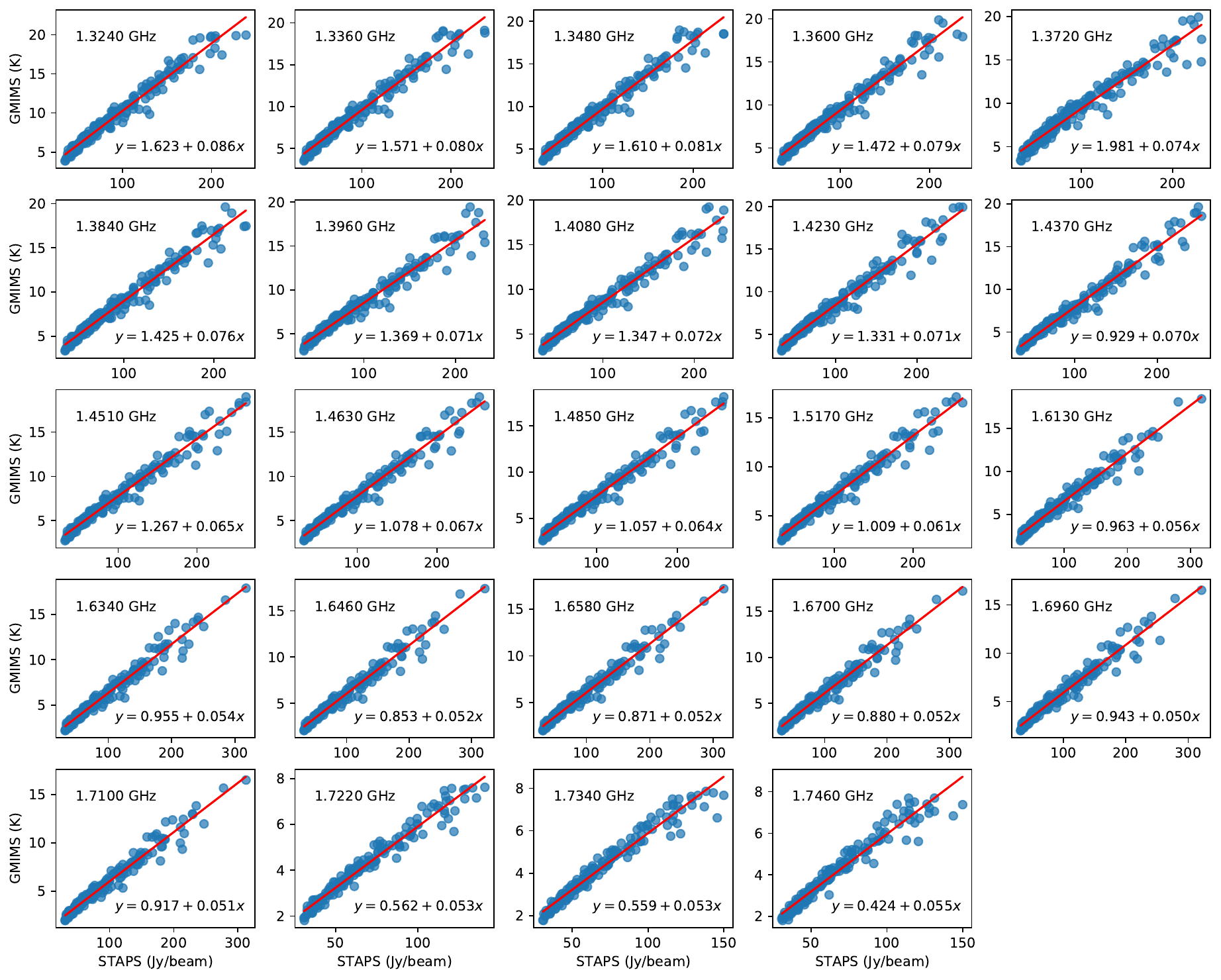}
    \caption{GMIMS pixel total intensity in K versus STAPS pixel total intensity in Jy beam$^{-1}$ for randomly selected frequencies between 1.3 and 1.8~GHz. The linear fits are also shown. }
    \label{fig:scales}
\end{figure*}
\end{appendix}
\end{document}